\pdfoutput=1
\documentclass[12pt,a4paper]{article}

\usepackage{ifthen} 
\newboolean{pdflatex}
\setboolean{pdflatex}{true} 
\usepackage[compat=1.1.0]{tikz-feynman}
\newboolean{articletitles}
\setboolean{articletitles}{true} 

\newboolean{uprightparticles}
\setboolean{uprightparticles}{false} 

\def\paperauthors{LHCb collaboration} 
\def\paperasciititle{Measurement of the Lund plane for light and beauty quark jets} 
\def\papertitle{Measurement of the Lund plane for light- and beauty-quark jets} 
\def\paperkeywords{{High Energy Physics}, {LHCb}} 
\def\papercopyright{\the\year\ CERN for the benefit of the LHCb collaboration} 
\def\paperlicence{CC BY 4.0 licence}
\def\paperlicenceurl{https://creativecommons.org/licenses/by/4.0/}

\newif\ifEnableSectionTOCLinks
\EnableSectionTOCLinksfalse 

\usepackage[top=1in, bottom=1.25in, left=1in, right=1in]{geometry}

\usepackage{microtype}
\usepackage{lineno}  
\usepackage{xspace} 
\usepackage{caption} 

\usepackage{graphicx}  
\usepackage{color}
\usepackage{colortbl}
\graphicspath{{./figs/}} 

\usepackage{amsmath} 
\usepackage{amssymb}
\usepackage{amsfonts}
\usepackage{upgreek} 

\newcommand*\patchAmsMathEnvironmentForLineno[1]{%
\expandafter\let\csname old#1\expandafter\endcsname\csname #1\endcsname
\expandafter\let\csname oldend#1\expandafter\endcsname\csname
end#1\endcsname
 \renewenvironment{#1}%
   {\linenomath\csname old#1\endcsname}%
   {\csname oldend#1\endcsname\endlinenomath}%
}
\newcommand*\patchBothAmsMathEnvironmentsForLineno[1]{%
  \patchAmsMathEnvironmentForLineno{#1}%
  \patchAmsMathEnvironmentForLineno{#1*}%
}
\AtBeginDocument{%
\patchBothAmsMathEnvironmentsForLineno{equation}%
\patchBothAmsMathEnvironmentsForLineno{align}%
\patchBothAmsMathEnvironmentsForLineno{flalign}%
\patchBothAmsMathEnvironmentsForLineno{alignat}%
\patchBothAmsMathEnvironmentsForLineno{gather}%
\patchBothAmsMathEnvironmentsForLineno{multline}%
\patchBothAmsMathEnvironmentsForLineno{eqnarray}%
}

\usepackage{hyperxmp}

\usepackage[pdftex,
            pdfauthor={\paperauthors},
            pdftitle={\paperasciititle},
            pdfkeywords={\paperkeywords},
            pdfcopyright={Copyright (C) \papercopyright},
            pdflicenseurl={\paperlicenceurl}]{hyperref}

\usepackage[colorinlistoftodos,textsize=scriptsize]{todonotes}

\usepackage[bottom,flushmargin,hang,multiple]{footmisc}

\usepackage[all]{hypcap} 

\usepackage{xspace} 
\usepackage{upgreek}

\def\lhcb   {\mbox{LHCb}\xspace}

\def\rich   {RICH\xspace}

\def\MagUp {\mbox{\em Mag\kern -0.05em Up}\xspace}

\ifthenelse{\boolean{uprightparticles}}%
{

 \def\Pmu         {\ensuremath{\upmu}\xspace}

 \def\Ppi         {\ensuremath{\uppi}\xspace}

 \def\Ppsi        {\ensuremath{\uppsi}\xspace}

 \def\PDelta      {\ensuremath{\Delta}\xspace}                 
 \def\PXi         {\ensuremath{\Xi}\xspace}                 
 \def\PLambda     {\ensuremath{\Lambda}\xspace}                 
 \def\PSigma      {\ensuremath{\Sigma}\xspace}                 
 \def\POmega      {\ensuremath{\Omega}\xspace}                 
 \def\PUpsilon    {\ensuremath{\Upsilon}\xspace}
 \let\oldPi\Pi
 \def\PPi         {\ensuremath{\oldPi}\xspace}

 \def\PB      {\ensuremath{\mathrm{B}}\xspace}                 
 \def\PD      {\ensuremath{\mathrm{D}}\xspace}                 
 \def\PJ      {\ensuremath{\mathrm{J}}\xspace}                 
 \def\PK      {\ensuremath{\mathrm{K}}\xspace}                 
 \def\PZ      {\ensuremath{\mathrm{Z}}\xspace}                 
 \def\Pb      {\ensuremath{\mathrm{b}}\xspace}                 
 \def\Pc      {\ensuremath{\mathrm{c}}\xspace}

 \def\Ps      {\ensuremath{\mathrm{s}}\xspace}

 \def\thebaroffset{0.0em}
}
{

 \def\Pmu         {\ensuremath{\mu}\xspace}

 \def\Ppi         {\ensuremath{\pi}\xspace}

 \def\Ppsi        {\ensuremath{\psi}\xspace}                 
                  
 \mathchardef\PDelta="7101
 \mathchardef\PXi="7104
 \mathchardef\PLambda="7103
 \mathchardef\PSigma="7106
 \mathchardef\POmega="710A
 \mathchardef\PUpsilon="7107
 \mathchardef\PPi="7105
 \def\PB      {\ensuremath{B}\xspace}                 
 \def\PD      {\ensuremath{D}\xspace}                 
 \def\PJ      {\ensuremath{J}\xspace}                 
 \def\PK      {\ensuremath{K}\xspace}                 
 \def\PZ      {\ensuremath{Z}\xspace}                 
 \def\Pb      {\ensuremath{b}\xspace}                 
 \def\Pc      {\ensuremath{c}\xspace}

 \def\Ps      {\ensuremath{s}\xspace}

 \def\thebaroffset{0.18em}
}
\newcommand{\offsetoverline}[2][\thebaroffset]{\kern #1\overline{\kern -#1 #2}}%

\makeatletter
\ifcase \@ptsize \relax
  \newcommand{\miniscule}{\@setfontsize\miniscule{4}{5}}
\or
  \newcommand{\miniscule}{\@setfontsize\miniscule{5}{6}}
\or
  \newcommand{\miniscule}{\@setfontsize\miniscule{5}{6}}
\fi
\makeatother

\DeclareRobustCommand{\optbar}[1]{\shortstack{{\miniscule (\rule[.5ex]{1.25em}{.18mm})}
  \\ [-.7ex] $#1$}}

\def\mup        {{\ensuremath{\Pmu^+}}\xspace}
\def\mun        {{\ensuremath{\Pmu^-}}\xspace} 
\def\mupm       {{\ensuremath{\Pmu^\pm}}\xspace}

\def\Z      {{\ensuremath{\PZ}}\xspace}

\def\squark    {{\ensuremath{\Ps}}\xspace}

\def\cquark    {{\ensuremath{\Pc}}\xspace}

\def\bquark    {{\ensuremath{\Pb}}\xspace}

\def\pion   {{\ensuremath{\Ppi}}\xspace}

\def\pipm   {{\ensuremath{\pion^\pm}}\xspace}

\def\kaon    {{\ensuremath{\PK}}\xspace}

\def\KorKbar {\kern \thebaroffset\optbar{\kern -\thebaroffset \PK}{}\xspace}

\def\Kpm     {{\ensuremath{\kaon^\pm}}\xspace}

\def\KL      {{\ensuremath{\kaon^0_{\mathrm{L}}}}\xspace}

\def\D       {{\ensuremath{\PD}}\xspace}

\def\DorDbar {\kern \thebaroffset\optbar{\kern -\thebaroffset \PD}\xspace}

\def\Dp      {{\ensuremath{\D^+}}\xspace}
\def\Dm      {{\ensuremath{\D^-}}\xspace}

\def\DpDm    {\ensuremath{\Dp {\kern -0.16em \Dm}}\xspace}

\def\B       {{\ensuremath{\PB}}\xspace}

\def\BorBbar {\kern \thebaroffset\optbar{\kern -\thebaroffset \PB}\xspace}

\def\Bd      {{\ensuremath{\B^0}}\xspace}

\def\BdorBdbar {\kern \thebaroffset\optbar{\kern -\thebaroffset \Bd}\xspace}
\def\Bu      {{\ensuremath{\B^+}}\xspace}

\def\Bpm     {{\ensuremath{\B^\pm}}\xspace}

\def\Bs      {{\ensuremath{\B^0_\squark}}\xspace}

\def\BsorBsbar {\kern \thebaroffset\optbar{\kern -\thebaroffset \Bs}\xspace}

\def\jpsi     {{\ensuremath{{\PJ\mskip -3mu/\mskip -2mu\Ppsi}}}\xspace}

\def\Y#1S{\ensuremath{\PUpsilon{(#1S)}}\xspace}

\def\Lz          {{\ensuremath{\PLambda}}\xspace}

\def\LorLbar     {\kern \thebaroffset\optbar{\kern -\thebaroffset \PLambda}\xspace}

\def\Lc          {{\ensuremath{\Lz^+_\cquark}}\xspace}

\def\to                 {\ensuremath{\rightarrow}\xspace}

\def\AT#1     {\ensuremath{A_{\mathrm{T}}^{#1}}\xspace}           

\def\C#1      {\ensuremath{\mathcal{C}_{#1}}\xspace}                       
\def\Cp#1     {\ensuremath{\mathcal{C}_{#1}^{'}}\xspace}                    
\def\Ceff#1   {\ensuremath{\mathcal{C}_{#1}^{\mathrm{(eff)}}}\xspace}        
\def\Cpeff#1  {\ensuremath{\mathcal{C}_{#1}^{'\mathrm{(eff)}}}\xspace}       
\def\Ope#1    {\ensuremath{\mathcal{O}_{#1}}\xspace}                       
\def\Opep#1   {\ensuremath{\mathcal{O}_{#1}^{'}}\xspace}                    

\newcommand{\nospaceunit}[1]{\ensuremath{\text{#1}}}       
\newcommand{\aunit}[1]{\ensuremath{\text{\,#1}}}       

\newcommand{\tev}{\aunit{Te\kern -0.1em V}\xspace}
\newcommand{\gev}{\aunit{Ge\kern -0.1em V}\xspace}
\newcommand{\mev}{\aunit{Me\kern -0.1em V}\xspace}
\newcommand{\kev}{\aunit{ke\kern -0.1em V}\xspace}
\newcommand{\ev}{\aunit{e\kern -0.1em V}\xspace}
 
\newcommand{\mevc}{\ensuremath{\aunit{Me\kern -0.1em V\!/}c}\xspace}
\newcommand{\gevc}{\ensuremath{\aunit{Ge\kern -0.1em V\!/}c}\xspace}
\newcommand{\mevcc}{\ensuremath{\aunit{Me\kern -0.1em V\!/}c^2}\xspace}
\newcommand{\gevcc}{\ensuremath{\aunit{Ge\kern -0.1em V\!/}c^2}\xspace}

\def\mum  {\ensuremath{\,\upmu\nospaceunit{m}}\xspace}

\def\fb   {\ensuremath{\aunit{fb}}\xspace}
\def\invfb   {\ensuremath{\fb^{-1}}\xspace}

\def\gsim{{~\raise.15em\hbox{$>$}\kern-.85em
          \lower.35em\hbox{$\sim$}~}\xspace}
\def\lsim{{~\raise.15em\hbox{$<$}\kern-.85em
          \lower.35em\hbox{$\sim$}~}\xspace}

\def\pt         {\ensuremath{p_{\mathrm{T}}}\xspace}

\def\evtgen     {\mbox{\textsc{EvtGen}}\xspace}

\def\geant      {\mbox{\textsc{Geant4}}\xspace}

\def\photos     {\mbox{\textsc{Photos}}\xspace}

\def\pythia     {\mbox{\textsc{Pythia}}\xspace}

\def\tell1  {TELL1\xspace}
\def\ukl1   {UKL1\xspace}

\newcommand{\ie}{\mbox{\itshape i.e.}\xspace}

\newcommand{\etc}{\mbox{\itshape etc.}\xspace}

\newcommand{\lhcborcid}[1]{\href{https://orcid.org/#1}{\hspace*{0.1em}\raisebox{-0.45ex}{\includegraphics[width=1em]{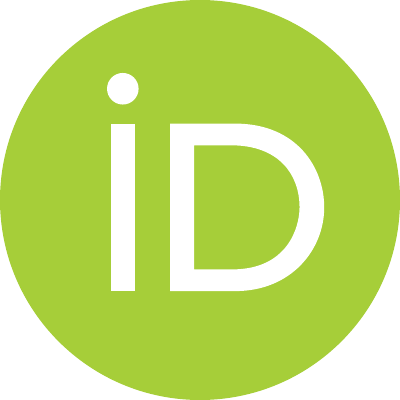}}}}

\hypersetup{
  colorlinks   = true, 
  urlcolor     = blue, 
  linkcolor    = blue, 
  citecolor    = red   
}

\ifEnableSectionTOCLinks
    \usepackage[explicit]{titlesec} 
    
    \let\oldcontentsline\contentsline
    \renewcommand

    \titleformat{\section}{\normalfont\Large\bf}{\hyperlink{tocsection.\thesection}{{\thesection} \parbox[t]{\dimexpr\textwidth-1pc}{#1}}}{1pc}{}

    \titleformat{\subsection}{\normalfont\bf}{\hyperlink{tocsubsection.\thesubsection}{{\thesubsection} \parbox[t]{\dimexpr\textwidth-1pc}{#1}}}{1pc}{}

    \titleformat{name=\section,numberless}[display]{}{}{0pt}{\normalfont\Huge\bfseries #1}
\fi

\usepackage{cite} 
\usepackage{mciteplus}

\usepackage{longtable} 
\usepackage{tabularx}

\def\lndR     {\ensuremath{\ln{\left(R/\Delta R\right)}}\xspace}

\def\lnkt     {\ensuremath{\ln{\left(k_{\mathrm{T}}/(\gevc)\right)}}\xspace}

\def\lnz    {\ensuremath{\ln{\left(1/z\right)}}\xspace}

\def\kt     {\ensuremath{k_\mathrm{T}}\xspace}
\def\yjet     {\ensuremath{y_\mathrm{jet}}\xspace}

\def\ptjet     {\ensuremath{p_\mathrm{T, jet}}\xspace}

\def\ptB     {\ensuremath{p_{\mathrm{T},\Bpm}}\xspace}

\def\Dzero      {\ensuremath{D^{0}}\xspace}
\def\Bu      {\ensuremath{B^{\pm}}\xspace}
\def\Zboson      {\ensuremath{Z}\xspace}

\def\Djet    {\ensuremath{D^{0}}-tagged jet\xspace}
\def\Bjet    {\ensuremath{B^{\pm}}-tagged\xspace jet}
\def\Zjet    {\ensuremath{Z}-tagged jet\xspace}

\def\Djets    {\ensuremath{D^{0}}-tagged jets\xspace}
\def\Bjets    {\ensuremath{B^{\pm}}-tagged  jets \xspace}
\def\Zjets    {\ensuremath{Z}-tagged jets \xspace}

\begin{document}

\renewcommand{\thefootnote}{\fnsymbol{footnote}}
\setcounter{footnote}{1}


\begin{titlepage}
\pagenumbering{roman}

\vspace*{-1.5cm}
\centerline{\large EUROPEAN ORGANIZATION FOR NUCLEAR RESEARCH (CERN)}
\vspace*{1.5cm}
\noindent
\begin{tabular*}{\linewidth}{lc@{\extracolsep{\fill}}r@{\extracolsep{0pt}}}
\ifthenelse{\boolean{pdflatex}}
{\vspace*{-1.5cm}\mbox{\!\!\!\includegraphics[width=.14\textwidth]{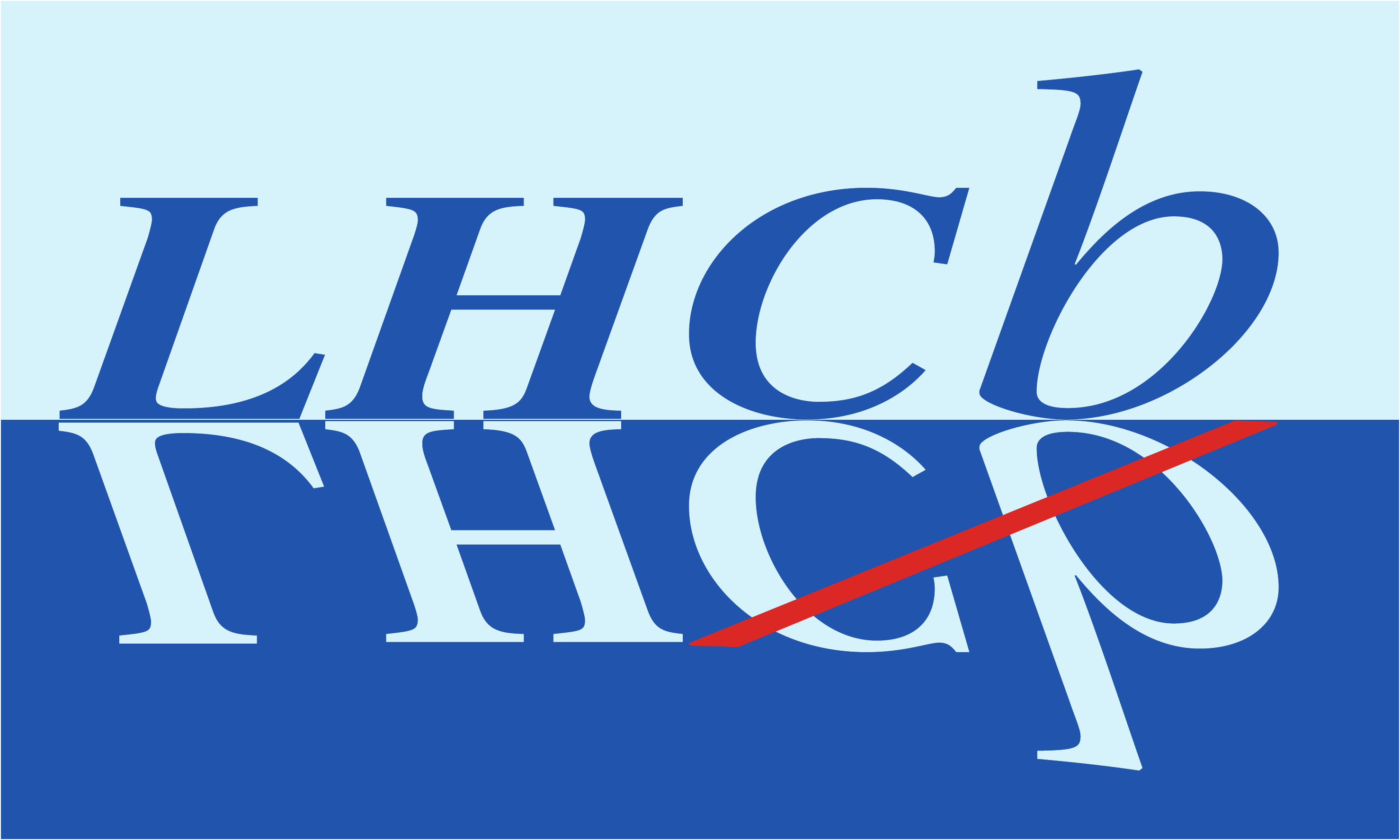}} & &}%
{\vspace*{-1.2cm}\mbox{\!\!\!\includegraphics[width=.12\textwidth]{figs/lhcb-logo.eps}} & &}%
\\
 & & CERN-EP-2025-093 \\  
 & & LHCb-PAPER-2025-010 \\  
 & & Oct 31, 2025 \\ 
 & & \\
\end{tabular*}

\vspace*{4.0cm}

{\normalfont\bfseries\boldmath\huge
\begin{center}
  \papertitle 
\end{center}
}

\vspace*{2.0cm}

\begin{center}
\paperauthors\footnote{Authors are listed at the end of this paper.}
\end{center}

\vspace{\fill}

\begin{abstract}
  \noindent
The substructure of jets in quantum chromodynamics (QCD) has garnered significant attention with the advent of infrared- and collinear-safe clustering algorithms and observables. A key question emerging from these studies is how in-jet emissions at soft and hard energy scales, across collinear and wide angles relative to the emitter, differ with the mass of the emitting parton. The Lund jet plane (LJP) is a perturbatively well-defined substructure observable that maps the radiation pattern of jets onto a plane, visually distinguishing emissions with different kinematic properties. Comparing LJP for jets containing hadrons of low versus high mass enables the testing of QCD splitting functions from first-principles calculations across both soft and hard regimes and at different radiation angles. 
 This article presents the first measurement of the LJP for light-quark-enriched and beauty-initiated jets at center-of-mass energy of 13\tev at LHCb. This marks the first direct observation of the dead-cone effect in beauty-quark jets, measured in the collinear region of the LJP.
\end{abstract}

\vspace*{2.0cm}

\begin{center}
  Published in Phys. Rev. D112 (2025) 072015
\end{center}

\vspace{\fill}

{\footnotesize 
\centerline{\copyright~\papercopyright. \href{\paperlicenceurl}{\paperlicence}.}}
\vspace*{2mm}

\end{titlepage}


\newpage
\setcounter{page}{2}
\mbox{~}
%
%
%
%


\renewcommand{\thefootnote}{\arabic{footnote}}
\setcounter{footnote}{0}


\cleardoublepage


\pagestyle{plain} 
\setcounter{page}{1}
\pagenumbering{arabic}



\def\lndR     {\ensuremath{\ln{\left(R/\Delta R\right)}}\xspace}

\def\lnkt     {\ensuremath{\ln{\left(k_{\mathrm{T}}/(\gevc)\right)}}\xspace}

\def\lnz    {\ensuremath{\ln{\left(1/z\right)}}\xspace}

\def\kt     {\ensuremath{k_\mathrm{T}}\xspace}
\def\yjet     {\ensuremath{y_\mathrm{jet}}\xspace}

\def\ptjet     {\ensuremath{p_\mathrm{T, jet}}\xspace}

\def\ptB     {\ensuremath{p_{\mathrm{T},\Bpm}}\xspace}

\def\Dzero      {\ensuremath{D^{0}}\xspace}
\def\Bu      {\ensuremath{B^{\pm}}\xspace}
\def\Zboson      {\ensuremath{Z}\xspace}

\def\Djet    {\ensuremath{D^{0}}-tagged jet\xspace}
\def\Bjet    {\ensuremath{B^{\pm}}-tagged\xspace jet}
\def\Zjet    {\ensuremath{Z}-tagged jet\xspace}

\def\Djets    {\ensuremath{D^{0}}-tagged jets\xspace}
\def\Bjets    {\ensuremath{B^{\pm}}-tagged  jets \xspace}
\def\Zjets    {\ensuremath{Z}-tagged jets \xspace}
\section{Introduction}
\label{sec:Introduction}

Due to color confinement in quantum chromodynamics (QCD), quarks and gluons are not directly accessible. The hard scattering of quarks and gluons is instead detected through boosted sprays of particles called jets. During the formation of a jet, a high-energy parton will radiate other partons leading to a cascade of QCD emissions called the parton shower. The initiating parton will lose energy until the hadronization scale is reached where partons combine into color-neutral bound states called hadrons. Heavy quarks exhibit hard fragmentation, transferring on average about 60\% (80\%) of their energy to the charm (beauty) hadron~\cite{DELPHI:1992pnf, OPAL:1994cct, OPAL:1995rqo, SLD:1999cuj, ALEPH:2001pfo}, whereas light quarks (up, down, and strange) radiate away most of their energy during the parton shower. These differences are fundamentally due to the dead-cone effect in QCD~\cite{Dokshitzer:1991fd}, which is the suppression of collinear radiation from massive quarks. The angle below which radiation is expected to be suppressed is referred to as the dead-cone angle ($\theta_q$) and increases with larger quark mass,
\begin{equation}
\label{eq:deadconeangle}
    \theta_{q} = \frac{m_{q}}{E_q}, 
\end{equation}
where $q$ represents a quark, $m_q$ is its mass, and $E_q$ is its energy. This effect is not limited to heavy quarks, but the low mass of light quarks reduces the dead cone to an immeasurably small angle. An indirect observation of the dead-cone effect was made by the DELPHI Collaboration at the Large Electron–Positron collider (LEP)~\cite{Battaglia:2004coa}, which observed a suppression in the angular distribution of fragmentation particles in beauty-quark jets relative to charm- and light-quark jets. On the other hand, a direct observation of the dead-cone effect was made only recently by the ALICE Collaboration in charm-initiated jets relative to inclusive jets~\cite{ALICE:2021aqk}. The direct observation required careful reconstruction of the branching history of the jet, taking advantage of the angular ordering property of QCD radiation~\cite{Cunqueiro:2018jbh}. 

By analyzing the properties of each emission in the branching history, a powerful representation of QCD radiation, called the Lund jet plane (LJP)~\cite{Andersson:1988gp,Dreyer:2018nbf}, can be constructed. The LJP is a jet substructure observable that separates the phase space of emissions in jets into different regions. Measuring the LJP is highly valuable for testing all-order logarithmic resummations in QCD and the parton shower and hadronization models of Monte Carlo event generators~\cite{Dreyer:2018nbf, Lifson:2020gua}.  The LJP has also been used for quark/gluon jet discrimination~\cite{Dreyer:2021hhr}, jet tagging\cite{Dreyer:2018nbf, Dreyer:2020brq, Khosa:2021cyk}, dark sector searches\cite{Cohen:2023mya}, obtaining gluon-enriched emissions~\cite{Baldenegro:2024pfb}, and more. Measuring the LJP for heavy-flavor (HF) jets provides an additional test of HF jet substructure calculations, where multiple scales (the jet transverse momentum, jet radius, and heavy-quark mass) are present which can spoil the convergence of the perturbative expansion\cite{Dhani:2024gtx}.

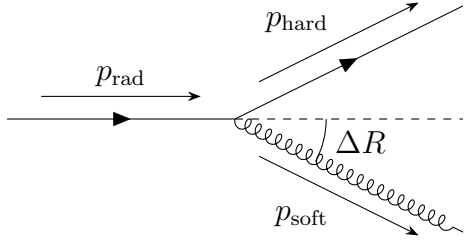
\begin{figure}[!t]
    \centering
\begin{tikzpicture}[scale=1.5]
\centering
    \begin{feynman}
        \vertex (a) at (-2,0);
        \vertex (b) at (0,0);
        \vertex (c) at (2,1);
        \vertex (d) at (2,-1);
        
        \diagram* {
            (a) -- [fermion, momentum={$p_{\text{rad}}$}] (b),
            (b) -- [fermion, momentum={$p_{\text{hard}}$}] (c),
            (b) -- [gluon, momentum'={$p_{\text{soft}}$}] (d)
        };
        
        \draw[dashed] (b) -- (2,0);
        
        \node at (1.1,-0.2) {\(\Delta R\)};

        \draw (0.8,0) arc[start angle=0, end angle=-28, radius=0.8];
    \end{feynman}
\end{tikzpicture}
\caption{An example Feynman diagram of a $1\to2$ parton splitting during the jet shower, in which a gluon is radiated from a quark.}
    \label{fig:splitting}
\end{figure}

To construct the LJP, it is helpful to first characterize the $1\to 2$ parton splitting in a jet, shown in Fig.~\ref{fig:splitting}. The four-momenta of the two decay partons are labeled by $p_{\mathrm{hard}}$ and $p_{\mathrm{soft}}$, defined according to the momentum transverse to the beam: \mbox{$p_{{\mathrm{T,hard}}} > p_{{\mathrm{T,soft}}}$}. The four-momentum of the radiating parton is given by $p_{\mathrm{rad}} = p_{\mathrm{hard}} + p_{\mathrm{soft}}$. The kinematics of the splitting are described using two representations: $\left[z, \Delta R\right]$ and $\left[\kt, \Delta R\right]$, where $z$ is the \pt fraction carried by the softer particle, \kt is the relative transverse momentum of the softer particle to the radiating parton, and $\Delta R$ is the angular distance in rapidity ($y$) and azimuthal angle ($\phi$) between the radiating parton and the soft emission:
\begin{equation}
\begin{aligned}
z &\equiv \frac{p_\mathrm{T, soft}}{p_\mathrm{T, soft} + p_\mathrm{T, hard}}\,, \\
\kt &\equiv p_\mathrm{T, soft} \,\, \Delta R \,,\\
\Delta R &\equiv \sqrt{\left(y_\mathrm{soft} - y_\mathrm{rad}\right)^2 + \left(\phi_\mathrm{soft} - \phi_\mathrm{rad}\right)^2} \,.
\end{aligned}
\end{equation}

The structure of the splitting function of parton $i$ to parton plus gluon $ig$ in QCD~\cite{Gribov:1972ri, Dokshitzer:1977sg, Altarelli:1977zs} in the soft-collinear limit is given by
\begin{equation}
    \mathrm{d}P_{i\to ig} \simeq \frac{2 \alpha_s(\kt) C_i}{\pi} \frac{\mathrm{d}\theta}{\theta} \frac{\mathrm{d}z}{z},
\end{equation}
where $\alpha_s(\kt)$ is the strong coupling constant at a momentum scale \kt, $C_i$ is the Casimir color factor of a quark ($C_F = 4/3$) or gluon ($C_A = 3$), and $\theta$ is the angle between the two product partons in the laboratory frame. Using the boost-invariant quantities along the beam direction in their logarithmic forms leads to a uniform distribution of soft-collinear emissions in the LJP:
\begin{equation}
    \mathrm{d}P_{i\to ig} \propto \mathrm{d}\left(\ln{\frac{1}{\Delta R}}\right) \mathrm{d}\left(\ln \frac{1}{z} \right).
\end{equation}

\begin{figure}[t]
    \centering
    \includegraphics[scale=0.37, page=1]{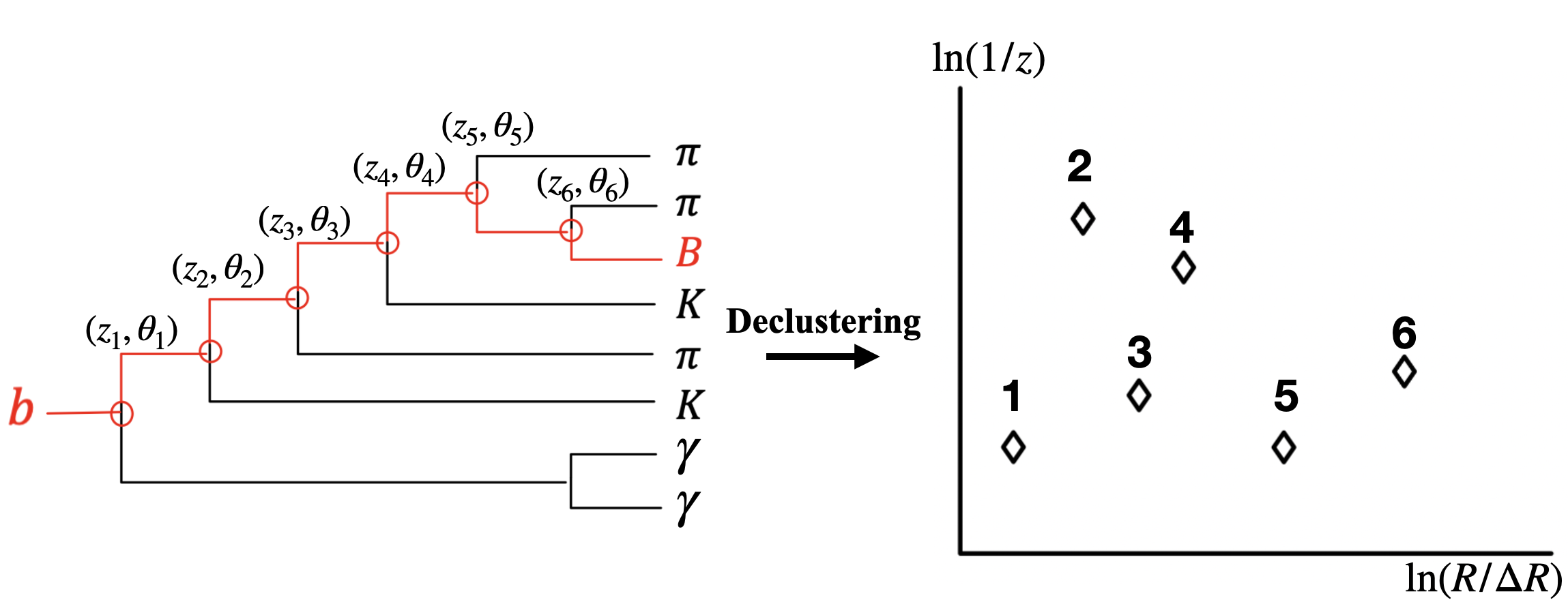}
    \caption{Left: example of the declustering procedure of reclustered C/A jet constituents following the branch containing a reconstructed $B$ meson, highlighted in red. Right: populating the LJP with emissions from the declustering procedure.}
    \label{fig:declustering}
\end{figure}

To populate the LJP, an iterative declustering procedure is performed by traversing the Cambridge/Aachen (C/A) jet tree~\cite{Dreyer:2018nbf} starting from its initiating parton. The C/A algorithm recombines particles that are close in angle first, following the approximate angular ordering of QCD radiation~\cite{Dokshitzer:1997in, Wobisch:1998wt}. The kinematic properties \kt, $z$, and $\Delta R$ of each consecutive emission are recorded, and the LJP is populated using these values, as shown in Fig.~\ref{fig:declustering}. At each declustering step, the branch of the harder particle is followed for light-quark and gluon jets, while for HF jets the branch containing a fully reconstructed beauty hadron is followed instead~\cite{Cunqueiro:2018jbh}. 

\begin{figure}[tp]
        \centering
        \includegraphics[scale=0.4]{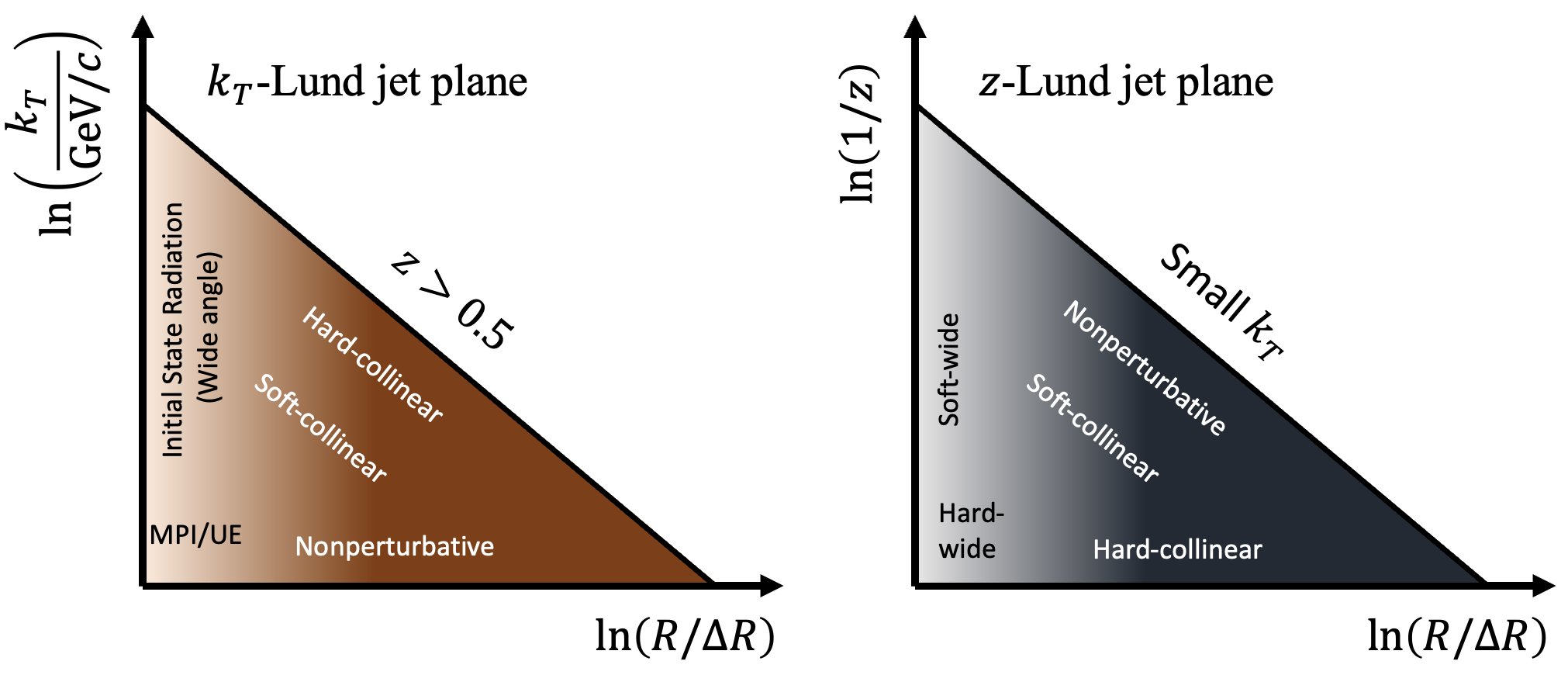}
    \caption{A sketch of the (left) \kt-Lund jet plane and (right) $z$-Lund jet plane showing the various types of gluon radiation in a jet, such as soft-collinear, nonperturbative, multiparton interactions (MPI) and underlying event (UE), \etc The color gradient indicates a suppression of emissions for darker colors due to the dead-cone effect in heavy-quark jets.}
    \label{fig:LJP_sketch}
\end{figure}

The LJP observables are then the emission densities $\rho$ given by
\begin{equation}
\label{eq:LJPkt}
    \rho(\kt, \Delta R) = \frac{1}{N_{\text{jets}}} \frac{\mathrm{d^2}n}{\mathrm{d}\lnkt \,\mathrm{d}\lndR}
\end{equation}
and 
\begin{equation}
\label{eq:LJPz}
    \rho(z, \Delta R) = \frac{1}{N_{\text{jets}}} \frac{\mathrm{d^2}n}{\mathrm{d}\lnz \,\mathrm{d}\lndR}\,,
\end{equation}
where $N_{\text{jets}}$ is the number of jets, $n$ is the number of emissions, and $R$ is the jet radius parameter. Equation~\ref{eq:LJPkt} is referred to as the \kt-LJP and Eq.~\ref{eq:LJPz} as the $z$-LJP. The \mbox{\kt-LJP} is useful for understanding the momentum scale of an emission, whether it is more perturbative or nonperturbative, and is structured primarily according to the running of $\alpha_s(\kt)$. The $z$-LJP is useful for accessing the behavior of the splitting functions in terms of the dimensionless quantity $z$ and separates soft and hard emissions in the LJP more cleanly. Figure~\ref{fig:LJP_sketch} shows sketches of the \kt-LJP and $z$-LJP with various types of gluon radiation in a jet.

There have been previous measurements of both the \kt and $z$ representations of the LJP. The ATLAS experiment has measured the $z$-LJP for charged tracks in inclusive high-\pt jets ($\pt > 675\gevc)$~\cite{ATLAS:2020bbn}, whereas the CMS experiment has measured the \kt-LJP for both small and large jet radii ($R = 0.4$ and $R = 0.8$)~\cite{CMS:2023lpp}. 
These measurements have focused on the LJP of midrapidity inclusive jets at the LHC. However, there have been no measurements so far of the LJP for jets initiated by heavy quarks, nor predominantly light-quark-initiated jets. The comparison of light- and heavy-quark radiation is complicated by the presence of gluon-initiated jets when using an inclusive sample. This is due to the larger Casimir color factor for gluons compared with quarks. Thus, a light-quark-enriched jet sample will facilitate direct comparisons of light- and heavy-quark radiation patterns in jets.

This paper presents measurements of the \kt-LJP and $z$-LJP in light-quark-enriched and beauty-quark jets in proton-proton ($pp$) collisions at a center-of-mass energy $\sqrt{s} = 13$\tev at the \lhcb experiment. The light-quark-enriched sample is obtained from jets recoiling off a \Z boson reconstructed through the dimuon channel \mbox{$\Z \to \mup \mun$}, referred to as $Z+$jets, which have previously been studied extensively by the \lhcb Collaboration~\cite{LHCb-PAPER-2013-058, LHCb-PAPER-2019-012, LHCb-PAPER-2022-013}. The sample of jets containing a beauty quark is obtained from jets built around fully reconstructed \Bpm mesons which are additionally required to pass a Winner-Take-All (WTA) flavor tag~\cite{Larkoski:2014uqa,Caletti:2022glq}, referred to as \Bpm-tagged jets. The \Bpm mesons are reconstructed through the \mbox{$\Bpm \to \jpsi(\to \mup \mun) \Kpm$} decay channel. A fully reconstructed HF hadron is used for three reasons: (1) the decay products of the HF hadron can contaminate the LJP, (2) the fully reconstructed HF hadron provides a unique branch connecting the root of the C/A tree to a final-state particle which can be followed during the declustering procedure (as in Fig.~\ref{fig:declustering}), and (3) the kinematics of the showering quark are better reconstructed through the parton shower. 

The data for this measurement are provided in the HEPData repository~\cite{HEPDATA}.


\section{Detector and data sample}
\label{sec:detector}
The description of the \lhcb detector presented here concerns the relevant data-taking period 2016--2018. The \lhcb detector~\cite{LHCb-DP-2008-001,LHCb-DP-2014-002} is a single-arm forward
spectrometer covering the \mbox{pseudorapidity} range $2 < \eta < 5$,
designed for the study of particles containing \bquark or \cquark
quarks. The detector includes a high-precision tracking system
consisting of a silicon-strip vertex detector surrounding the $pp$
interaction region~\cite{LHCb-DP-2014-001}, a large-area silicon-strip detector located upstream of a dipole magnet with a bending power of about
$4{\mathrm{\,T\,m}}$, and three stations of silicon-strip detectors and straw drift tubes~\cite{LHCb-DP-2017-001}
placed downstream of the magnet.
The tracking system provides a measurement of the momentum of charged particles with a relative uncertainty that varies from 0.5\% at low momentum to 1.0\% at 200\gevc~\cite{LHCb-DP-2014-002}.
The minimum distance of a track to a primary $pp$ collision vertex (PV), the impact parameter, is measured with a resolution of $(15+29/\pt)\mum$, where \pt is given in\,\gevc.
Different types of charged hadrons are distinguished using information
from two ring-imaging Cherenkov detectors (\rich)~\cite{LHCb-DP-2012-003}. 
Photons, electrons and hadrons are identified by a calorimeter system consisting of scintillating-pad and preshower detectors, an electromagnetic (ECAL) and a hadronic (HCAL) calorimeter. Muons are identified by a system composed of alternating layers of iron and multiwire proportional chambers~\cite{LHCB-DP-2012-002}.

The data samples used in this analysis correspond to an integrated luminosity of 5.4\invfb.
The online event selection is performed by a trigger~\cite{LHCb-DP-2012-004}, 
which consists of a hardware stage, based on information from the calorimeter and muon
systems, followed by a software stage, which applies a full event
reconstruction.

Simulation is used to model the detector acceptance and selection requirements, as well as to study potential backgrounds. Proton-proton collisions are generated using \pythia8~\cite{Sjostrand:2007gs} with a specific \lhcb configuration~\cite{LHCb-PROC-2010-056}, and decays of unstable particles are simulated by \evtgen~\cite{Lange:2001uf}, with final-state radiation generated by \photos~\cite{davidson2015photos}. The interaction of generated particles with the detector and its response is implemented using the \geant toolkit~\cite{Agostinelli:2002hh, Allison:2006ve} as detailed in Ref.~\cite{LHCb-PROC-2011-006}. To improve the simulated response of the RICH detectors, particle identification (PID) variables are corrected using calibration data from \mbox{$D^{*+} \to D^0\pi^+$} and \mbox{$D^0 \to K^-\pi^+$} decays for $K^\pm$ and $\pi^\pm$ mesons, as well as \mbox{$\Lc \to pK^-\pi^+$} decays for protons~\cite{LHCb-DP-2018-001}. This correction adjusts the variables to match the data distributions, accounting for track kinematics.


\section{Event reconstruction and selection}
\label{sec:eventrecoandsel}
In all of the data samples, only events with a single reconstructed PV are selected to mitigate the effects of pileup on jet reconstruction.

\subsection{\boldmath{\Z}-boson selection}
The reconstruction of $Z$-boson candidates is performed through the combination of two oppositely charged muons, at least one of which must pass the hardware-trigger and subsequent software-trigger requirements. The trigger requirements select good-quality tracks identified as muons with $\pt(\mupm)>12.5\gevc$. Further selections are then imposed on the muons, requiring them to have \mbox{$\pt(\mupm) > 20\gevc$}, \mbox{$2.0 < \eta(\mupm) < 4.5$}, and an invariant mass of \mbox{$60 < M_{\mu\mu} < 120\gevcc$}, as outlined in Refs.~\cite{LHCb-PAPER-2013-058, LHCb-PAPER-2019-012}. They must also meet the track reconstruction and muon identification criteria specified in Ref.~\cite{LHCb-PAPER-2021-037}. 

\subsection{Beauty-hadron selection}
\label{subsec:beautyselection}
The \mbox{$\Bpm \to \jpsi(\to \mup \mun) \Kpm$} signal candidates are first required
to pass the hardware trigger, which selects events with at least one $\mupm$ candidate with \mbox{$\pt(\mupm) > 1.35$--1.80\gevc},
or at least one \mup \mun candidate with \mbox{$\sqrt{\pt(\mup)\pt(\mun)} > 1.3$--1.5\gevc},
where the thresholds varied during the data taking period~\cite{LHCb-DP-2019-001}. In the subsequent software
trigger, the pair of muons is required to have an invariant mass above 2.7\gevcc.
The online event selection uses the pair of oppositely charged muons to reconstruct \jpsi-meson candidates that are required to be within 100\mevcc of the known mass~\cite{PDG2024}, displaced from the PV, and have a \pt greater than 2\gevc. Kaons are identified using information from the RICH detectors, which further suppresses \Bpm candidates arising from the Cabibbo-suppressed decay \mbox{$\Bpm \to \jpsi(\to \mup \mun) \pipm$}. They must also be  inconsistent with originating from the PV and have a transverse momentum greater than 250\mevc. The \Bpm candidates are required to have a mass greater than 5.15\gevcc and less than 5.58\gevcc. The asymmetry in the mass window around the nominal \Bpm-meson mass is chosen to remove partially reconstructed background from \mbox{$B \to \jpsi \left(K\pi\right)X$} decays, where $X$ represents unreconstructed decay products, which appears as a broad shoulder on the lower region of the mass spectrum. No minimum $\pt$ requirement is imposed on the \Bpm-meson candidates.

When performing fits to the \Bpm-meson mass spectrum, candidates are constrained to originate from the PV and the \jpsi-meson candidates are constrained to their known mass value, yielding a better resolution on the candidate mass peak~\cite{Hulsbergen:2005pu}. These fits are performed after jet reconstruction and are discussed in Sec.~\ref{subsec:jetreconstruction}.

\subsection{Jet reconstruction}
\label{subsec:jetreconstruction}
Jets are reconstructed offline using a particle flow algorithm~\cite{LHCb-PAPER-2013-058}, clustering neutral and charged candidates with \texttt{FastJet3.4.1}~\cite{Cacciari:2011ma} using the anti-\kt algorithm~\cite{Cacciari:2008gp} with $E$-scheme recombination, which defines the jet axis as the sum of the four-momenta of reconstructed jet constituents. The jets are reconstructed with a resolution (radius) parameter of $R=0.5$ and a minimum jet transverse momentum \mbox{$\ptjet > 20\gevc$} within \mbox{$2.5 < \yjet < 4.0$}. The rapidity range is restricted to ensure all particles inside the jet are within the acceptance and have good reconstruction efficiency. Since mass effects are significant, rapidity is used in calculating angular distances as opposed to pseudorapidity~\cite{Gallicchio:2018elxQuitUsingPseudo}. 

For the $Z+$jets sample, the jet must be produced in the opposite direction to the \Z-boson candidate, determined by an azimuthal separation \mbox{$|\Delta \phi_{Z,\text{jet}}|$} greater than $\frac{7\pi}{8}$, and is rejected if a muon is found within \mbox{$\Delta R_{\mu,\mathrm{jet}} = \sqrt{(\Delta y_{\mu,\mathrm{jet}})^2 + (\Delta \phi_{\mu,\mathrm{jet}})^2} < 0.5$} relative to the jet axis. Fake jets arising from noise or isolated energetic leptons are suppressed with jet identification requirements~\cite{LHCb-PAPER-2013-058}:
\begin{enumerate}
    \item jets are required to contain at least two particles matched to the same PV;
    \item jets are required to contain at least one track with $\pt > 1.2\gevc$;
    \item no single particle carries more than 75\% of \ptjet; and
    \item the fraction of \ptjet carried by charged particles must be greater than 10\%.
\end{enumerate} 
The jet with the highest \pt in the event, satisfying all the selection requirements, is chosen for further analysis. Around 13\,400 $Z$+jet candidates pass the selection requirements.

The \Bpm-tagged jets are reconstructed with some key differences. Prior to jet clustering, the decay products of the \Bpm meson are replaced with the \Bpm meson candidate, and after jet clustering, the jet whose constituents include the \Bpm meson is analyzed. These \Bpm jets are further required to be tagged using ``WTA flavor"~\cite{Caletti:2022glq}, which defines the flavor of a jet by the net flavor of the constituents that lie along the WTA axis~\cite{Larkoski:2014uqa}. In this case, the WTA axis is required to coincide with the direction of the \Bpm meson. 
The WTA axis is found by reclustering the jet using the C/A algorithm with the \pt recombination scheme, where the direction of the WTA axis at each recombination is determined by the direction of the higher-\pt branch, until all particles have been recombined. The WTA axis is insensitive to soft particles since it always lies along the direction of the harder particle during a recombination step~\cite{Larkoski:2014uqa}. The WTA-flavor tag ensures that theoretical calculations are infrared-safe to all orders, with a well understood dependence on the collinear unsafety~\cite{Caletti:2022glq}.

In addition to selection requirements on the jets as a whole, selection criteria on the jet constituents are imposed after jet clustering to select high-quality final-state particles. The momentum of each jet constituent is required to be greater than 4\gevc and the transverse momentum greater than 250\mevc which avoids low-efficiency regions in phase space. Charged jet constituents are required to have a good track-fit quality. 

Following jet reconstruction, the number of signal \Bpm-meson candidates is determined from fits to the mass distribution of all \Bpm-meson candidates in bins of \Bpm transverse momentum. The signal distribution is fitted using two double-sided Crystal Ball functions~\cite{Skwarnicki:1986xj} with shared peak position and tail parameters. The tail parameters are fixed from fits in simulation, while the peak position and widths are left as free parameters. The combinatorial background is fitted using a second-order polynomial function and its fraction under the signal peak decreases from 20\% for \mbox{$\ptB < 4\gevc$} to negligible values for \mbox{$\ptB > 35\gevc$}. The misidentified background arising from Cabibbo-suppressed \mbox{$\Bpm \to \jpsi(\to \mup \mun) \pipm$} decays is studied in simulation and is also fitted using a double-sided Crystal Ball function, but is found to be negligible in the overall fits. 

The \Bpm-mass signal region is defined by the intervals \mbox{$[m_0-N\sigma, m_0 + N\sigma]$} where $m_0$ is the peak position of the signal distribution, $\sigma$ is the smaller of the two widths (which provides better control over the mass window), and $N=2$ for \mbox{$\ptB < 5\gevc$}, $N=3$ for \mbox{$ 5 < \ptB <  8\gevc$}, and $N=4$ for \mbox{$\ptB > 8\gevc$}. The variable mass window is chosen due to the larger combinatorial background fraction at low \Bpm transverse momentum. The edges of the signal region are approximately \mbox{$\left[5.24, 5.31\right]\gevcc$} in most bins of \ptB.

The jet constituents are reclustered with the C/A algorithm using a modified version of the \texttt{LundGenerator} package in \texttt{fjcontrib 1.041}~\cite{Dreyer:2018nbf,Cacciari:2011ma}. The default package allows the declustering of the C/A tree following the hardest branch, and the modification allows the declustering of the C/A tree following the branch containing the HF hadron~\cite{Cunqueiro:2018jbh}, as shown in Fig.~\ref{fig:declustering}. Requiring the WTA-flavor tag on the \Bpm-tagged jets leads to the HF branch being the same as the hardest branch for over 99\% of splittings. The remaining $<1\%$ are due to the different recombination schemes between the LJP reconstruction and the WTA-flavor tag, which use the $E$-scheme and the WTA \pt-scheme, respectively. During declustering, spurious emissions arising from decays of excited \Bpm mesons are suppressed by requiring \mbox{$\lnkt > -3$}. These emissions manifest as a region of constant $\kt$ in the LJP below \mbox{$\lnkt= -3$} deep in the nonperturbative region.

\begin{figure}
    \centering
    \includegraphics[width=0.8\linewidth,page=3]{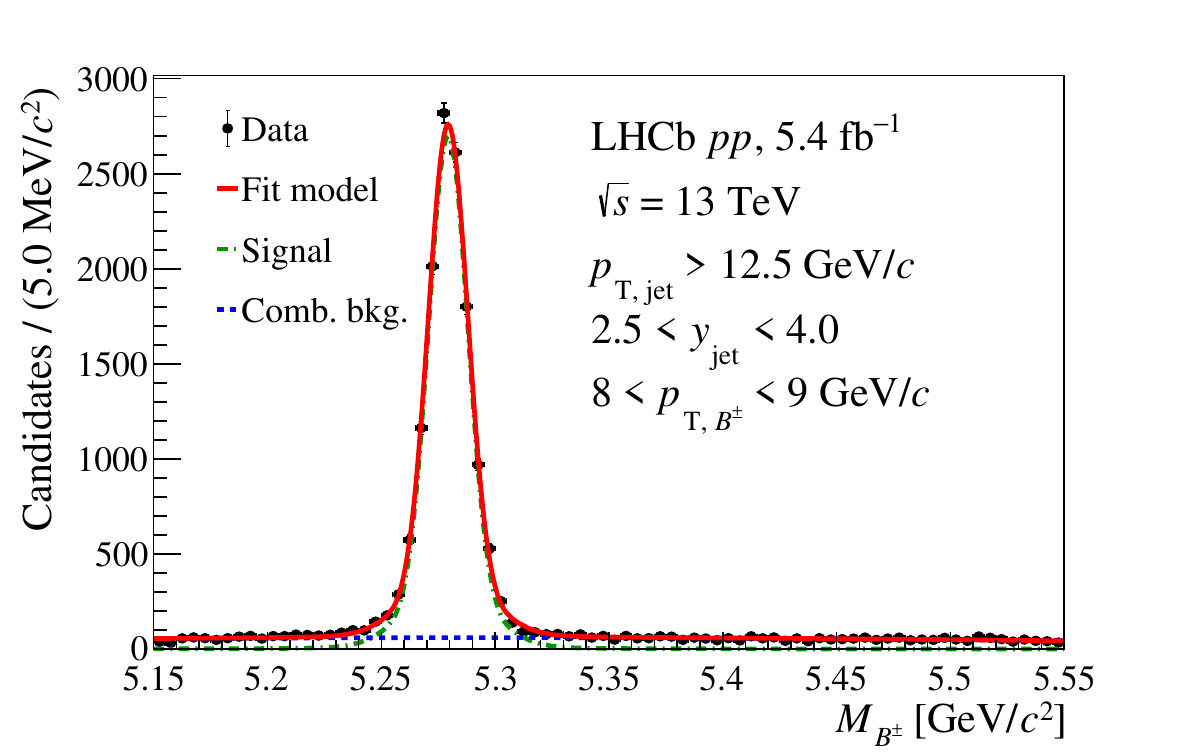}
    \caption{Mass distribution for \Bpm-meson candidates with  \mbox{$8 < p_{\mathrm{T},B^{\pm}} < 9$\gevc}. The fit model and its components are also shown.}
    \label{fig:examplemassfit}
\end{figure}

A sideband subtraction method is used to extract the LJP of signal candidates. The LJP from sidebands, dominated by combinatorial background, is scaled and subtracted from the LJP in the signal region, which contains both signal and background. Around 45\,300 signal \Bpm-tagged jet candidates with $\ptjet > 20\gevc$ are obtained from the fitting procedure. Figure~\ref{fig:examplemassfit} shows an example of the fits to the mass distribution for \Bpm-meson candidates.

\section{Selection efficiencies}
The detector performance and selection criteria discussed in Sec.~\ref{sec:detector} and \ref{sec:eventrecoandsel} lead to inefficiencies in reconstructing the event tag (\Z boson or \Bpm meson) and the jets. In this section, the corrections at the level of the event, including the event-tag reconstruction efficiency and the jet reconstruction efficiency 
are reported. These efficiencies are used to weight each event when constructing the LJP.

\subsection{\boldmath{\Z}+jet reconstruction efficiencies}
The efficiency of reconstructing $Z+$jets involves the efficiency of reconstructing the $\Z\to\mup\mun$ decay as well as the recoiling jet. The total efficiency of reconstructing the \mup\mun pair is given by the product of three types of muon efficiencies: the tracking efficiency, the muon identification efficiency, and the trigger efficiency. These efficiencies have been determined as a function of muon $\eta$ and \pt, using a tag-and-probe method as in previous \lhcb measurements~\cite{LHCb-DP-2013-002,LHCb-PAPER-2012-008, LHCb-PAPER-2015-001, LHCb-PAPER-2015-049, LHCb-PAPER-2014-033}.
The average muon tracking efficiency and muon identification efficiency of these high-\pt muons ($\pt > 20\gevc$) is about 95\%, and the average single-muon-trigger efficiency is about 80\%~\cite{LHCb-PAPER-2016-021}. The total trigger efficiency is larger than the single-muon-trigger efficiency at about 95\%, since either muon can trigger the event~\cite{LHCb-PAPER-2016-021}. The product of the trigger, tracking, and PID efficiencies of the muons yields the event-tag efficiency. 

The jet reconstruction efficiency is studied in simulation with \pythia 8~\cite{Sjostrand:2007gs} on samples of $Z+$jet events as a function of \ptjet. Fake jets can also contaminate the sample and must be corrected for by a purity factor. A detector-level jet (formed from reconstructed objects in the detector) is geometrically matched to a particle-level jet (formed from particles before the event enters the detector simulation) on an event-by-event basis if it has the smallest angular distance $\Delta R$ with the particle-level jet, satisfying $\Delta R < 0.5$. 

\begin{figure}[!tb]
    \centering
    \includegraphics[scale=0.6]{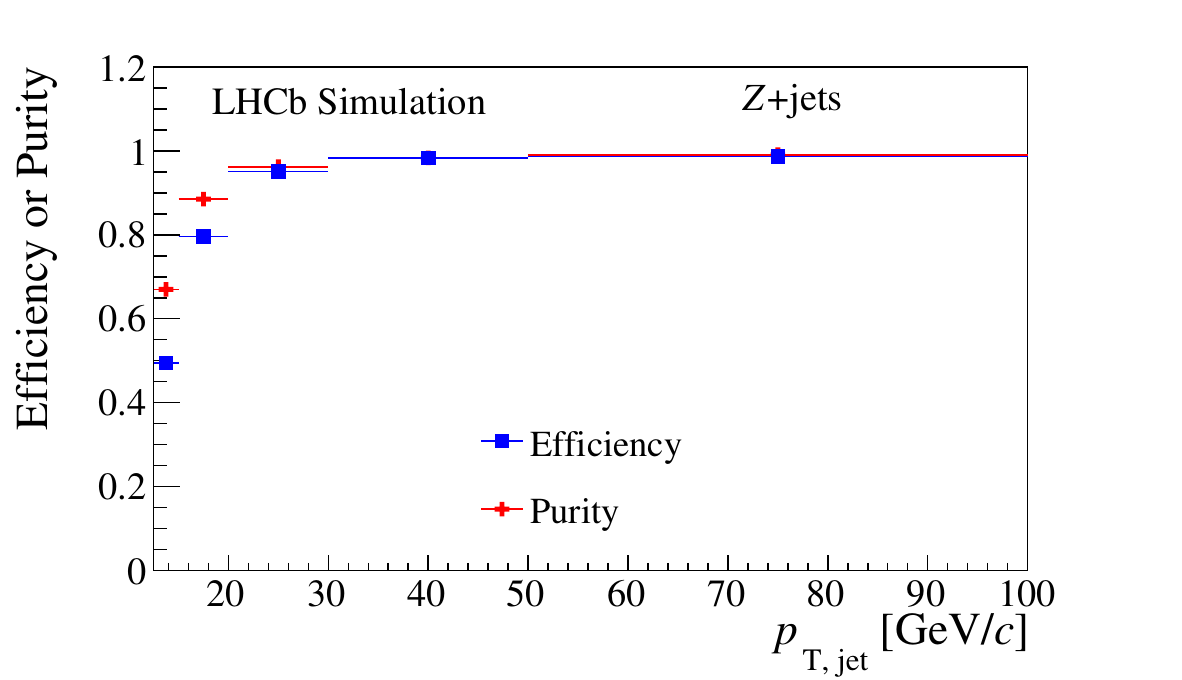}
    \caption{Jet reconstruction efficiency and purity of $Z+$jets as a function of particle-level \ptjet.}
    \label{fig:Zjet_jetrecoeff}
\end{figure}

Figure \ref{fig:Zjet_jetrecoeff} shows the jet reconstruction efficiency and purity in $Z+$jets as a function of particle-level \ptjet. The efficiency is 50\% for \mbox{$\ptjet < 15\gevc$} and increases to nearly 100\% for $\ptjet > 30\gevc$. The purity is reduced at low \ptjet due to fake-jet contamination or true jets that have \ptjet less than 12.5\gevc reconstructed above 12.5\gevc. The final measurement is performed on jets with \mbox{$\ptjet > 20\gevc$}, while the bins below 20\gevc are used to correct for bin migrations as described in Sec.~\ref{sec:LJPEfficienciesUnfolding}.

\subsection{\boldmath{\Bpm}-tagged jet reconstruction efficiency}
The efficiency of reconstructing a \Bpm-tagged jet involves the product of the efficiency of reconstructing the \Kpm meson and \mup \mun pair, the trigger efficiency, and the efficiency of reconstructing a jet containing the \Bpm candidate in the signal region passing the selection requirements. The total \Bpm-tagged jet reconstruction efficiency is computed in simulation and is then corrected by efficiencies obtained in data through calibration and control samples as follows. The total efficiency of reconstructing the \Kpm meson and \mup \mun pair is the product of the tracking and PID efficiencies. These are determined using the tag-and-probe method with control channels~\cite{LHCb-DP-2013-002, LHCb-DP-2018-001} mentioned in Sec.~\ref{sec:detector}. Both the tracking and PID efficiencies vary between 90\% and 100\% over a wide range of track momenta from 4\gevc to 200\gevc. The efficiency of the trigger selection is calculated using control channels in data~\cite{LHCb-PUB-2014-039}. The trigger efficiency increases from 60\% to 80\% as the \jpsi \pt increases from 2\gevc to 40\gevc. There is a weak dependence on the rapidity of the \jpsi meson which is also taken into account. 

After reconstructing the decay products of the \Bpm candidate, further selection criteria are imposed on the \jpsi and \Bpm mesons as discussed in Sec.~\ref{subsec:beautyselection}. These selections lead to a nontrivial dependence of the reconstruction efficiency of the \Bpm meson on its transverse momentum which is studied in simulation using samples of $\Bpm \to \jpsi \Kpm$ decays. 
The \Bpm-meson reconstruction efficiency, including the trigger, tracking, and PID efficiencies of its decay products, grows from 10\% to 30\% as \ptB increases from 2\gevc to 50\gevc. 

Once the \Bpm meson has been reconstructed, the remaining efficiency factors quantify the jet reconstruction and WTA-flavor tagging. These efficiencies are also studied in simulation as a function of \ptjet, \ptB, and the jet rapidity. A detector-level jet is matched to a \Bpm-tagged particle-level jet on an event-by-event basis if it contains a \Bpm meson and is within $\Delta R < 0.5$ from the particle-level jet. 

\begin{figure}
    \centering
    \begin{minipage}{0.49\textwidth}
        \centering
        \includegraphics[width=\linewidth]{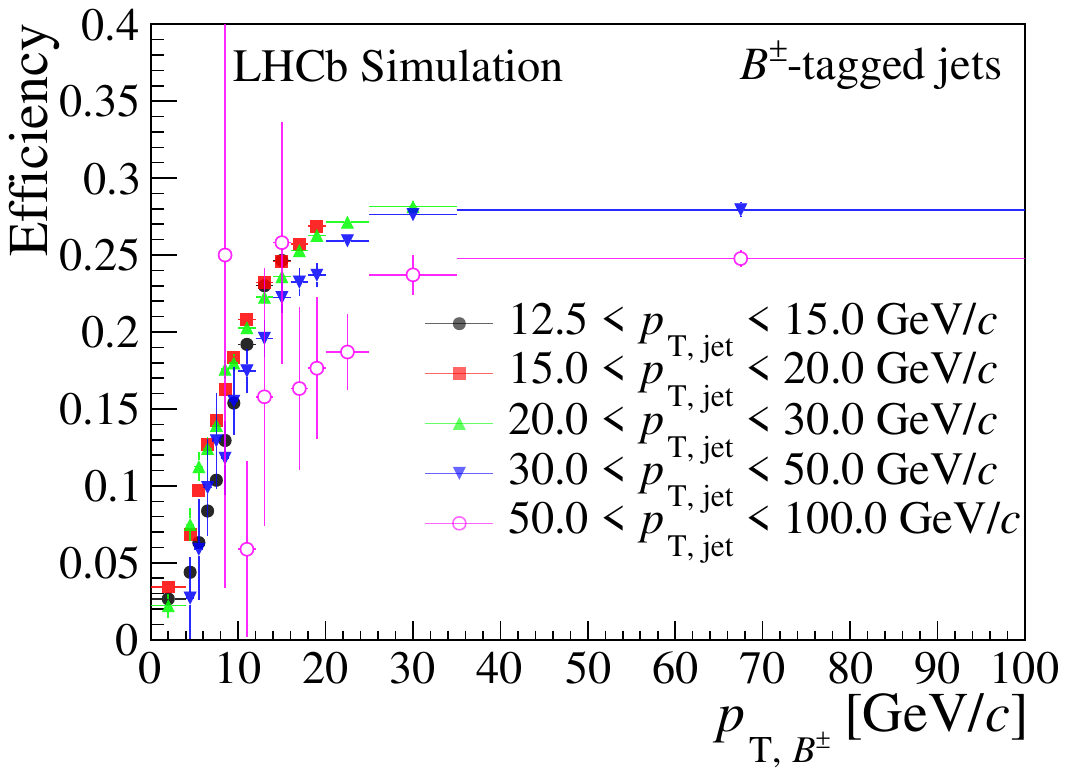}
    \end{minipage}
    \begin{minipage}{0.49\textwidth}
        \centering
        \includegraphics[width=\linewidth]{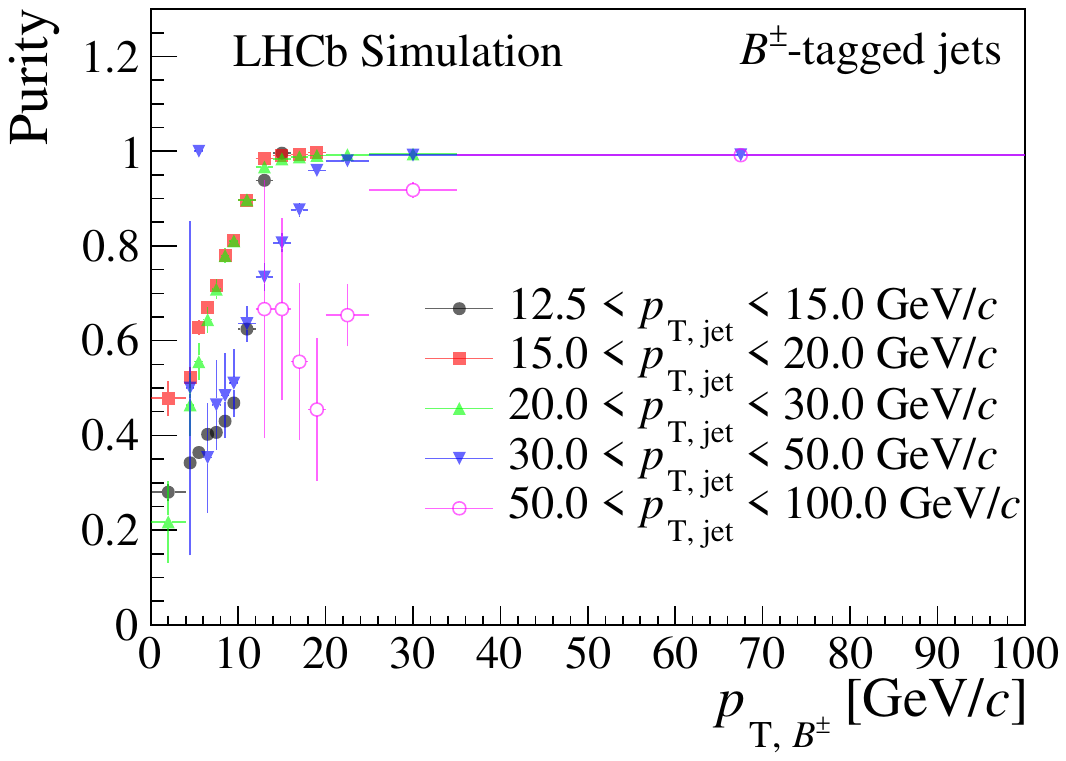}
    \end{minipage}
    \caption{The \Bpm-tagged jet (left) reconstruction efficiency and (right) purity as a function of \ptB and particle-level \ptjet. These values do not include corrections from calibration and control samples.}
    \label{fig:Bjetreceff_HFptjetpt}
\end{figure}

Figure \ref{fig:Bjetreceff_HFptjetpt} shows the \Bpm-tagged-jet reconstruction efficiency and purity as a function of \ptB and particle-level \ptjet in simulation with no corrections from the calibration and control samples applied, integrated over the jet rapidity for visualization purposes. The efficiency grows from about 3\% to 25\% across different \ptjet bins as \ptB increases from 2\gevc to 50\gevc. The purity in different \ptjet bins drops significantly for \ptB below 15\gevc due to the WTA-flavor tag which requires the \Bpm meson to be aligned with the WTA axis. The large uncertainties at low \ptB in the highest \ptjet bin contribute negligibly to the final LJP uncertainties due to the low overall contribution of these events to the total sample.

\section{Lund jet plane efficiencies and unfolding}
\label{sec:LJPEfficienciesUnfolding}
Due to inefficiencies in reconstructing final-state particles, the finite resolution of the \lhcb detector, and bin migrations into and from the measured \ptjet region, emissions in the LJP along with the jet \pt spectrum must be corrected to particle level. Although track momenta are measured with better than 1\% precision, tracking inefficiencies combined with the lower momentum resolution of neutral particles contribute to an overall smearing effect on the reconstructed jet kinematics and its substructure. To correct for these effects, D'Agostini's iterative Bayesian method~\cite{DAgostini:1994fjx}, implemented in \textsc{RooUnfold}~\cite{Adye:2011gm}, is used to unfold the measured distributions in data to particle level. The response matrix and emission reconstruction efficiency and purity are studied in  simulation. 

A six-dimensional response matrix in three variables---\ptjet and the two LJP axes---is used to unfold the LJPs for $Z$+jets and \Bpm-tagged jets. For $Z$+jets, the event-level efficiency maps are largely constant in the LJP, and so cancel out with the normalization $N_{\mathrm{jets}}$. For \Bpm-tagged jets, the event-level efficiency maps vary across the LJP, so, ideally an eight-dimensional response matrix in the four variables \ptjet, \ptB, and the two LJP axes would be constructed. However, due to statistical limitations in the simulation samples, as well as the limitation of \textsc{RooUnfold} in utilizing four variables, the unfolding is performed with the aforementioned six-dimensional response matrix. The \Bpm-tagged-jet reconstruction efficiency is applied prior to unfolding as an event weight. A systematic uncertainty due to this choice is estimated from closure tests (see Sec.~\ref{subsec:ljpunfolding}).

\subsection{Emission reconstruction efficiency}
\label{subsec:emissionreco}

To compute the emission reconstruction efficiencies and purities, a two-way unique matching procedure \cite{CMS:2023lpp} between particle-level and detector-level emissions in simulation is performed. Given a pair of matched jets, the declustering procedure applied to the particle(detector)-level jet produces a list of emissions at particle (detector) level. Starting at particle level, the list is iterated over from wide- to narrow-angle emissions finding the detector-level emission with the smallest angular distance satisfying $\Delta R < 0.1$. If such a detector-level emission is found, the matched emissions are removed from their respective lists, and the matching procedure is continued for the next particle-level emission until all particle-level emissions have been considered. This procedure is then repeated starting with the detector-level emissions and finding their matched particle-level emissions. A pair of emissions is two-way matched if it exists in both iterations. Any unmatched detector-level emissions contribute to the purity correction, while any unmatched particle-level emissions contribute to the efficiency correction. The two-way matched emissions are used to construct the response matrix which unfolds the LJP to particle level. The corrections and unfolding matrix are binned in \ptjet and the two LJP variables. 

The efficiency generally increases with more perturbative emissions [larger $\ln(\kt/(\text{GeV/}c))$] from 50\% to 80\% in the bulk of the \kt-LJP, and is highest for intermediate angles around $\lndR = 1$. The efficiency is lowest, reaching 30\%, for wide-angle low-\kt emissions. The purity follows a similar trend in the \kt-LJP.
These observations are similar between $Z+$jets and \Bpm-tagged jets. 
In the $z$-LJP, the efficiency is larger for harder emissions [smaller $\ln(1/z)$], increasing from 50\% to 80\% and is again highest for intermediate angles around $\lndR = 1$. Harder emissions have higher energy and are more likely to be reconstructed. On the other hand, collinear emissions might be mismatched during the two-way matching procedure due to the angular resolution and thus decrease the reconstruction efficiency in the collinear region [large $\ln(R/\Delta R)$]. Wide-angle emissions at the jet boundary might not be reconstructed due to the final-state particles ending up outside of the jet. 

The purity also increases with harder emissions, and is largest for intermediate angles around $\lndR = 1$. Particles outside of the jet at particle level might end up in the jet at detector level, contributing to wide-angle emissions and lowering the purity. The relatively low purity in the very collinear region is again due to the mismatch of collinear emissions in the two-way matching procedure. These observations are similar between $Z+$jets and \Bpm-tagged jets.

\subsection{Lund jet plane response matrix and unfolding}
\label{subsec:ljpunfolding}
A six-dimensional response matrix in \ptjet and the two LJP variables is built from matched emissions in simulation. For the \kt-LJP, this involves unfolding in $\left[\ptjet, \lnkt, \lndR\right]$, while for the $z$-LJP, the unfolding involves the variables $\left[\ptjet, \lnz, \lndR\right]$.

The final LJP distributions are constructed as follows: starting with the weighted LJPs, the distributions are multiplied by the emission reconstruction purity, unfolded, then divided by the emission reconstruction efficiency. The LJPs are then normalized by the number of jets $N_{\mathrm{jets}}$ obtained from a separate one-dimensional unfolding of the jet \pt spectrum. 

This procedure is validated through a closure test in simulation. The detector-level LJP distribution is smeared, bin-by-bin, by the statistical uncertainties of the LJP distribution in data, then is subject to the purity correction, unfolding, and the efficiency correction. The smeared and corrected detector-level LJP is then divided by the particle-level distribution. This procedure is repeated 30 times, and the average absolute deviation from unity in each bin is taken as a nonclosure systematic uncertainty discussed in Sec.~\ref{sec:systematics}.

The closure tests are also used to determine the number of unfolding iterations. The number is chosen such that the average absolute difference between the unfolded detector-level distribution and the particle-level distribution stabilizes. This is found to occur at four unfolding iterations. A systematic uncertainty arising from this choice is determined in Sec.~\ref{sec:systematics}. Statistical correlations between bins are found to be negligible and are not considered in the final results.


\section{Systematic uncertainties}
\label{sec:systematics}

Several sources of systematic uncertainties are identified in this measurement. Unless specified, the separate systematic uncertainties are assumed to be uncorrelated and are added in quadrature to obtain the total systematic uncertainty. The systematic uncertainty for each variation is found by repeating the analysis with this variation and taking the difference between the baseline and varied distributions. 

The systematic uncertainty associated with the signal extraction and signal region determination of \Bpm-tagged jets is estimated by refitting the mass distribution using a Student's $t$-distribution for the signal function. The effect is found to be negligible. The background contribution to the $Z+$jets sample is negligible, so this systematic uncertainty is not considered. To determine the systematic uncertainty associated with the sideband subtraction procedure, the sideband regions are varied in two ways: (1) the lower edge of the lower sideband is increased by $1\sigma$ and the higher edge of the upper sideband is decreased by $1\sigma$ and (2) the higher edge of the lower sideband is decreased by $1\sigma$ and the lower edge of the upper sideband is increased by $1\sigma$. The systematic uncertainty from each variation is averaged by taking the root mean square value. The overall systematic uncertainty is found to vary from $< 1\%$ in the bulk to 10\% at the edges of the LJPs.

The systematic uncertainty associated with reconstructing and identifying the decay products of the event tag is considered. For $Z+$jets, the uncertainties on the tracking and PID efficiencies of high-\pt muons are found to have a negligible effect on the LJPs. For the \Bpm-tagged jets, the main contributions to the uncertainty in the tracking efficiency are the statistical uncertainty of the calibration sample and the uncertainty on the material budget which affects how particles---particularly kaons---interact with the detector and are tracked~\cite{LHCb-DP-2018-001}. Propagated to the LJPs, these systematic uncertainties are less than 1\% in most bins. The systematic uncertainty associated with the \Bpm-meson selection is estimated by tightening the selection criteria and repeating the analysis. The resulting systematic uncertainty is negligible in the bulk of the LJPs and grows to 10\% at the edges.

The uncertainty on the trigger efficiency for high-\pt muons in the $Z+$jets sample is found to be negligible. 
For the \Bpm-tagged jet events, the uncertainty on the trigger efficiency is less than 1\% in most bins of the  LJPs. The systematic uncertainties on the jet energy scale and jet energy resolution are found by comparing the \pt balance between the \Z boson and the recoiling jet, $\ptjet/p_{\mathrm{T},Z}$, in simulation to that in data using $Z+$jet events~\cite{LHCb-PAPER-2013-058, LHCb-PAPER-2019-012, LHCb-PAPER-2022-013}. These uncertainties have been found to be about 3\% for the jet energy scale and 11\% for the jet energy resolution for jets with $\ptjet > 20\gev/c$~\cite{LHCb-PAPER-2022-013}. The uncertainty on the jet energy scale is propagated to the LJP by shifting \ptjet by the uncertainty and repeating the unfolding chain. For the \Bpm-tagged jets, only the non-\Bpm component of the jet~\cite{LHCb-PAPER-2016-064} is shifted since the four-momentum of the \Bpm meson is well reconstructed. The uncertainty on the jet energy resolution is propagated to the LJP by smearing the jet \pt and repeating the unfolding chain. Again, for \Bpm-tagged jets, only the non-\Bpm component of \ptjet is smeared. These contributions to the total systematic uncertainty are most significant at the edges of the LJP, increasing to 10\%.

The systematic uncertainty on the jet identification requirements applied to the $Z+$jets sample and described in Sec.~\ref{subsec:jetreconstruction} is estimated by tightening the selection and repeating the analysis.
The systematic uncertainty is found to be negligible. The number of tracks in the jet is considered as a jet identification selection on the \Bpm-tagged jets and a systematic uncertainty is estimated by varying the selection from at least one track (the jet must contain the \Bpm-meson) to at least two tracks (\ie at least one other track in addition to the \Bpm-meson). This uncertainty is about 1\% in the bulk of the \Bpm-tagged LJP and increases to about 4\% at the edges.

There are three main sources of uncertainty related to the unfolding procedure: the nonclosure uncertainty, the response matrix prior, and the number of iterations. The nonclosure systematic uncertainty is estimated from the closure test performed in Sec.~\ref{subsec:ljpunfolding}. This uncertainty takes into account any limitations inherent to the unfolding procedure in addition to any systematic uncertainty due to the choice of using a six-dimensional response matrix as opposed to an eight-dimensional response matrix as discussed in Sec.~\ref{sec:LJPEfficienciesUnfolding}. While unfolding does not explicitly depend on the weighting of the prior, the \textsc{RooUnfold} implementation of the Bayesian procedure uses a projection of this matrix to determine the prior. The unfolding prior is nominally based on \pythia8 with \lhcb tune. Here, a different prior is tested by reweighting the response matrix by two factors: an emission-level weight related to the jet substructure, and an event-level weight related to \ptjet and $p_{\mathrm{T},Z}$ or $\ptB$. The emission weight is computed by taking the ratio between detector-level simulation and data of a three-dimensional LJP in $\left[\lnz, \lnkt, \lndR\right]$. The event weight is the ratio of the joint $[\ptB, \ptjet, \yjet]$ distribution at detector level to data. After reweighting the response matrix using these ratios, the analysis chain is repeated using the reweighted response matrices (and thus prior distributions), and the difference between the modified and baseline result is taken as the systematic uncertainty. To estimate the uncertainty associated with the number of iterations, the unfolding is performed with one more and one less iteration in data. The three different uncertainties associated with the unfolding are treated as independent and are added in quadrature to obtain the final unfolding systematic uncertainty. The unfolding systematic uncertainty---and in particular the nonclosure and unfolding prior uncertainty---is the dominant uncertainty across the LJP. It varies from 0\% in the bulk to 20\% at the edges of the $Z$+jet LJPs, and from 2\% in the bulk to 50\% at the edges of the \Bpm-tagged LJPs.

Differences between data and simulation in the efficiency of reconstructing a jet constituent contribute a systematic uncertainty on the LJP. This systematic uncertainty is propagated to the LJP by removing jet constituents in simulation with a probability given by the value of this systematic uncertainty (in \%) depending on the charge of the constituent. The uncertainty on the track-reconstruction efficiency including hadronic interactions is about 1.5\%~\cite{LHCb-DP-2013-002} assuming that most tracks are charged pions. The systematic uncertainty on the ECAL reconstruction efficiency of photons and $\pi^0$ mesons has been studied in \Bpm-meson decays\cite{Govorkova:2015vqa} and is found to be around 4\%. There is no similar study available for the HCAL reconstruction efficiency, so as a first approximation, particles reconstructed in the HCAL are also dropped with a 4\% probability. The uncertainty on the final-state particle reconstruction efficiency is found to grow from $<1\%$ in the bulk to about 14\% in the edges of the LJP.

An upper bound on the systematic uncertainty due to the uncertainty in the HCAL reconstruction efficiency is found by looking at the ratio in simulation of the particle-level LJP without neutrons and \KL mesons clustered in jets to the distribution with these particles included. This is equivalent to removing the HCAL at detector level. It is found that even after removing these particles entirely, the uncertainties are contained between 0 and 10\% in the bulk of the LJP, increasing to 40\% at the edges. This shows that the true uncertainty will be much lower than the dominant unfolding systematic uncertainty.

The matching procedure outlined in Sec.~\ref{subsec:emissionreco} can be performed by reversing the baseline ordering and iterating from narrowest to widest splittings. Following this reverse ordering, the response matrices, emission reconstruction efficiencies, and purities are recalculated and applied to the data. The resulting systematic uncertainty is negligible.

The dominant systematic uncertainty in this analysis is due to the unfolding. The range of each source of systematic uncertainty is summarized in Tables \ref{tab:zjetsystematic} and \ref{tab:bjetsystematic} for the $Z+$jets and \Bpm-tagged jets, respectively. These ranges exclude any sparsely populated bins at the edges of the LJP.
\begin{table}[!tb]
    \caption{Summary of systematic uncertainties for the $Z+$jets LJPs. These ranges exclude any sparsely populated bins at the edges of the Lund jet planes.}
    \label{tab:zjetsystematic}
    \centering
    \begin{tabular}{lccc}  
         Source&$\kt$-LJP uncertainty (\%) &$z$-LJP uncertainty (\%)\\ \hline  
         Tracking efficiency&  \textless\,\,1&\textless\,\,1\\   
         PID efficiency&  \textless\,\,1&\textless\,\,1\\  
         Trigger efficiency&  \textless\,\,1&\textless\,\,1\\  
         Jet energy scale&  0 -- 10& 0 -- 10\\   
         Jet energy resolution& 0 -- 10& 0 -- 5 \,\\ 
         Jet identification& 0 -- 2 \,& 0 -- 2 \,\\ 
         Unfolding prior& 0 -- 13& 0 -- 15\\ 
         Unfolding nonclosure& 2 -- 10 & 2 -- 20\\ 
         Unfolding iterations& 0 -- 5 \,& 0 -- 5 \,\\ 
         Jet-constituent-\\reconstruction efficiency& 0 -- 14&0 -- 10\\   
 \hline
 Total& 0 -- 23&0 -- 25\\ \hline 
    \end{tabular}
\end{table}

\begin{table}[!ttb]
    \caption{Summary of systematic uncertainties for the \Bpm-tagged jet LJPs. These ranges exclude any sparsely populated bins at the edges of the Lund jet planes.}
    \label{tab:bjetsystematic}
    \centering
    \begin{tabular}{lccc}  
         Source& $\kt$-LJP uncertainty (\%) &$z$-LJP uncertainty (\%)\\ \hline  
         Background subtraction& 0 -- 10 &0 -- 11\\ 
         Tracking efficiency&  \textless\,\,1&\textless\,\,1\\   
         PID efficiency&  \textless\,\,1&\textless\,\,1\\  
         Trigger efficiency&  \textless\,\,1&\textless\,\,1\\  
         Jet energy scale&0 -- 8 \,&0 -- 6 \,\\   
         Jet energy resolution& 0 -- 14 &0 -- 10\\ 
         Jet identification& 0 -- 5 \,&0 -- 7 \,\\ 
         Unfolding prior& 0 -- 25&0 -- 50\\ 
         Unfolding nonclosure& 3 -- 50&3 -- 55\\ 
         Unfolding iterations& 0 -- 6 \,&0 -- 9 \,\\   
         Reconstruction and selection& 0 -- 10&0 -- 10\\   
         Jet-constituent-\\reconstruction efficiency& 0 -- 6 \,& 0 -- 6 \,\\ 
\hline
 Total&$0 - 60$ &$0 - 65$\\ \hline 
    \end{tabular}
\end{table}

\section{Results}
\label{sec:finalresults}

\begin{figure}[!ttb]
    \centering
\begin{minipage}[b]{0.49\textwidth}
\includegraphics[width=\textwidth]{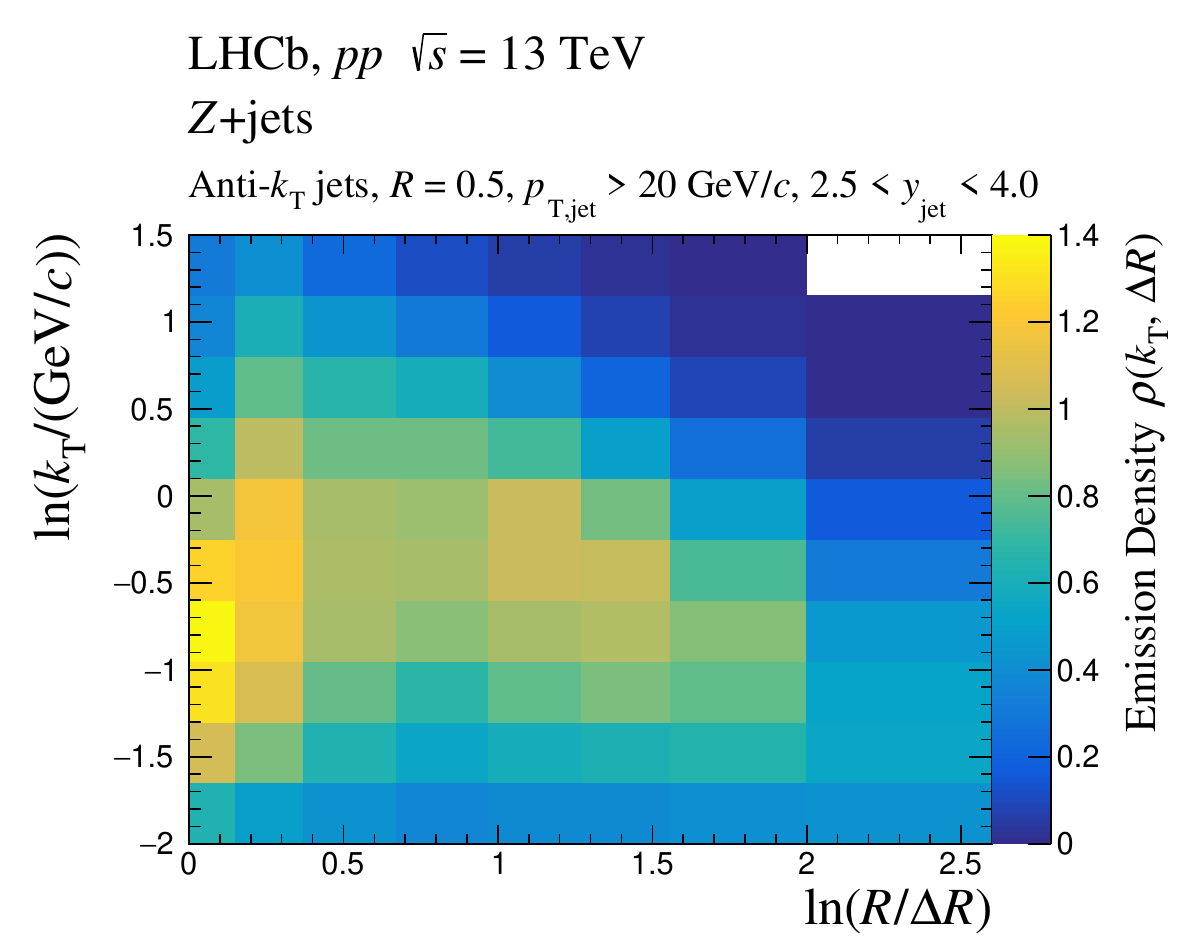}
        \end{minipage}
        \hfill
        \centering
\begin{minipage}[b]{0.49\textwidth}
\includegraphics[width=\textwidth]{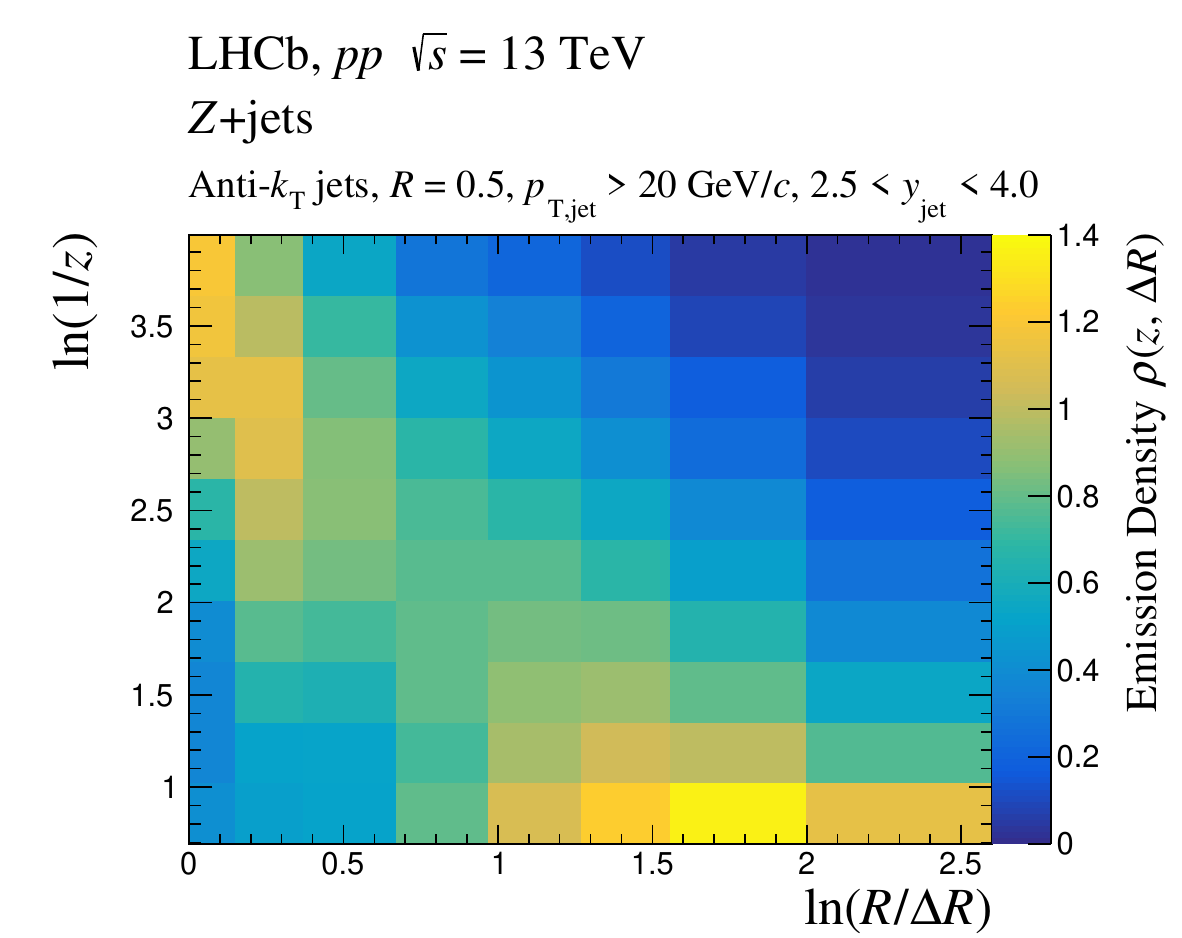}
        \end{minipage}
        \hfill
        \centering
\begin{minipage}[b]{0.49\textwidth}
\includegraphics[width=\textwidth]{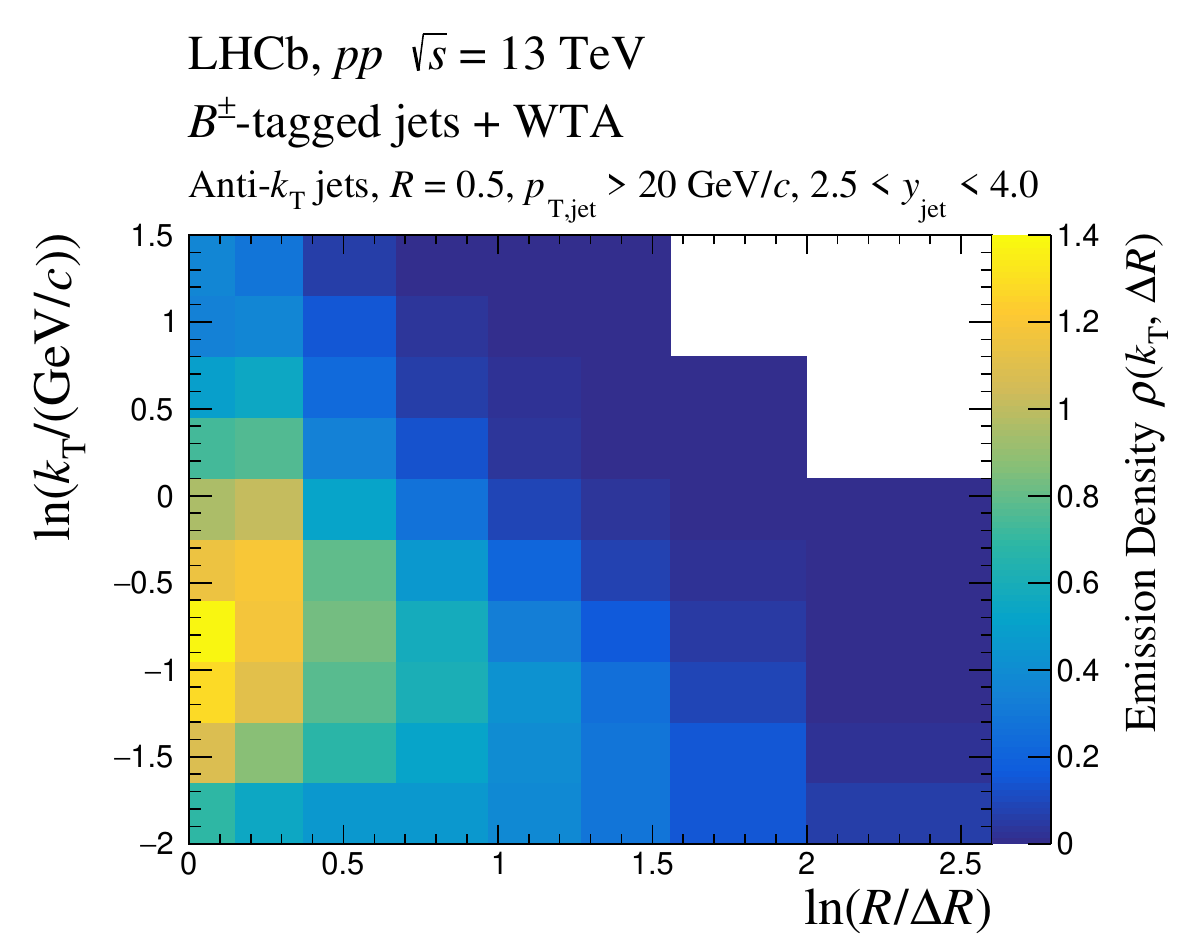}
        \end{minipage}
        \hfill
        \centering
\begin{minipage}[b]{0.49\textwidth}
\includegraphics[width=\textwidth]{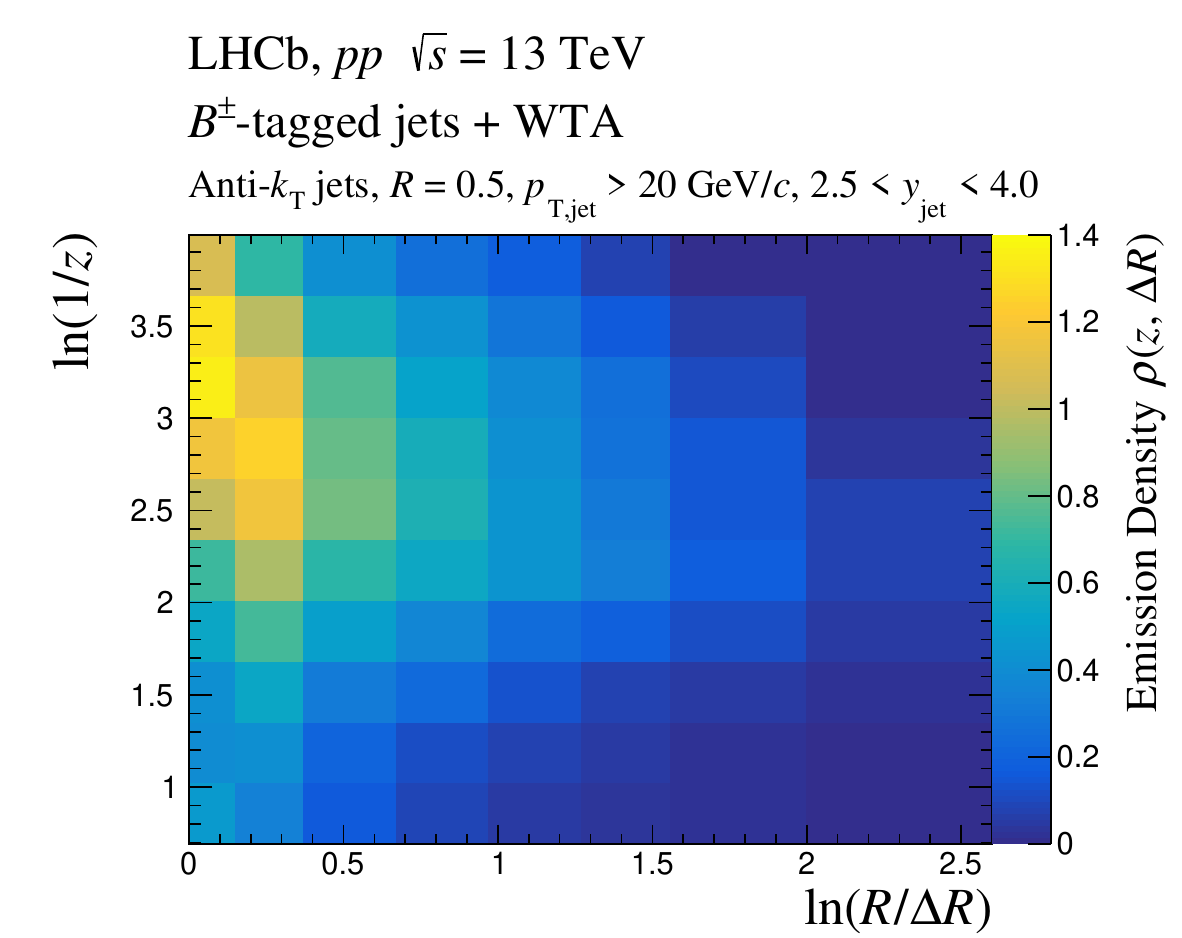}
        \end{minipage}
        \centering
\caption{Measurements of the (top left) \kt-LJP and (top right) $z$-LJP for $Z$+jets, and the (bottom left) \kt-LJP and (bottom right) $z$-LJP for \Bpm-tagged jets in $pp$ collisions at $\sqrt{s} = 13\tev$ at forward rapidity. The beauty dead-cone effect is observed in the collinear region where radiation is strongly suppressed in the \Bpm-tagged jets LJPs relative to the $Z$+jets LJPs.}
    \label{fig:2DLJPs}

\end{figure}

The following distributions are measured with $R=0.5$ anti-$k_{\mathrm{T}}$ jets reclustered with the C/A algorithm in the kinematic region $\ptjet>20\gevc$, and $2.5<\yjet<4.0$ at $\sqrt{s} = 13\tev$. Figure \ref{fig:2DLJPs} shows the unfolded measurements of the LJP for $Z$+jets \kt-LJP, $Z$+jets $z$-LJP, \Bpm-tagged jets \kt-LJP, and \Bpm-tagged jets $z$-LJP in data. From the \kt-LJP of $Z+$jets (top left), the running of the strong coupling $\alpha_s(\kt)$ in the \kt-LJP is readily identified by noting the decrease in emission density in the direction of large \kt (more perturbative), and the rise in emission density in the direction of lower \kt towards the confinement scale $\Lambda_{\mathrm{QCD}}$, until a turnaround is seen in the nonperturbative confinement region. The region where \mbox{$0.5 < \lndR < 1.0$} and \mbox{$-1.0 < \lnkt < 0.5$} shows an approximately constant density of emissions which is identified as the soft-collinear region. Likewise in the $z$-LJP of $Z+$jets (top right), various splitting regions can be readily identified. The top left region exhibits a high density of emissions corresponding to the wide-angle soft radiation. Similarly, the bottom right region also exhibits a high density of emissions corresponding to the hard-collinear emissions. The transition region along the diagonal is uniformly populated with emissions, indicating the soft-collinear region. 

Stark differences are seen in the collinear region \mbox{($\lndR > 0.5$ or $\Delta R < 0.3$)} in the \kt-LJP between the \Bpm-tagged jets (bottom left) and the $Z+$jets (top left), where the emission density is significantly lower in the heavy quark case. This is direct evidence of the dead-cone effect in HF jets. The wide-angle region on the other hand, where mass effects are not expected, shows similar behavior between the $Z$+jets and \Bpm-tagged jets. Likewise for the $z$-LJP, stark differences between the \Bpm-tagged jets (bottom right) and the $Z+$jets (top right) are seen in the collinear region, where the hard-collinear radiation almost completely vanishes for the \Bpm-tagged jets. This is again direct evidence of the dead-cone effect in heavy-quark jets.

\begin{figure*}[t]
    \centering
    \begin{minipage}[b]{0.49\textwidth}
        \includegraphics[width=\textwidth]{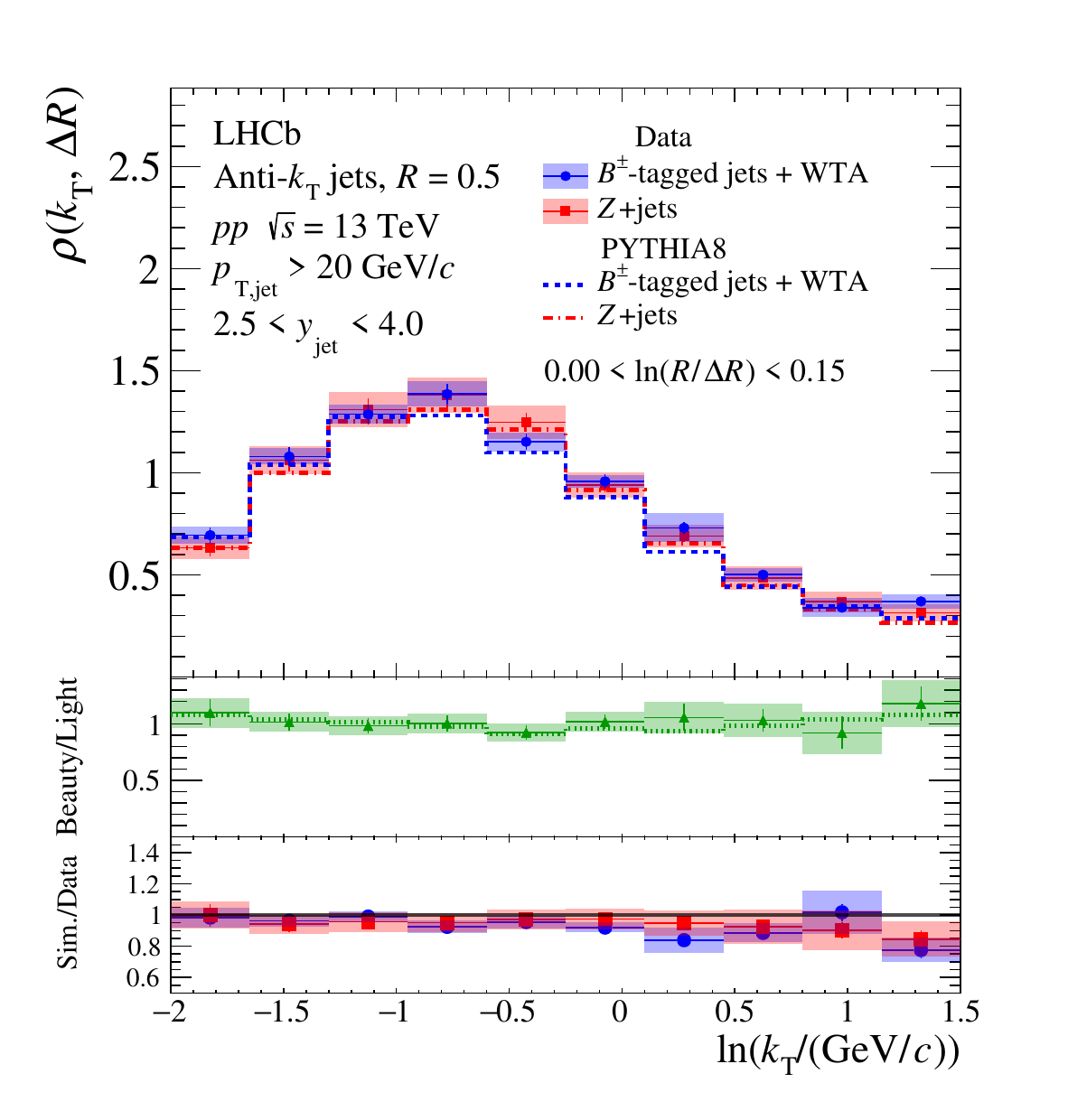}
    \end{minipage}
    \hfill
    \begin{minipage}[b]{0.49\textwidth}
        \includegraphics[width=\textwidth]{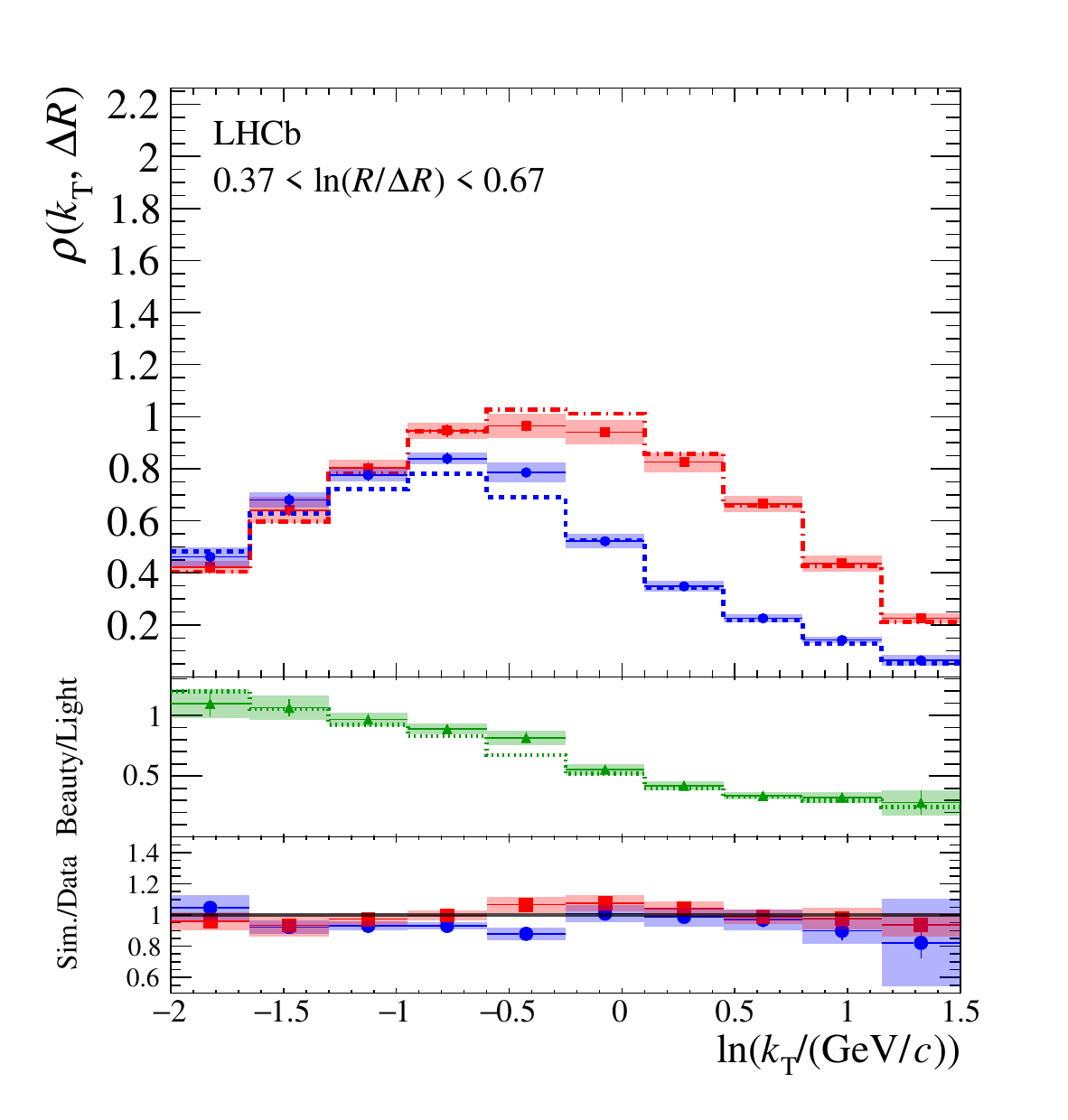}
    \end{minipage}
    \begin{minipage}[b]{0.49\textwidth}
        \includegraphics[width=\textwidth]{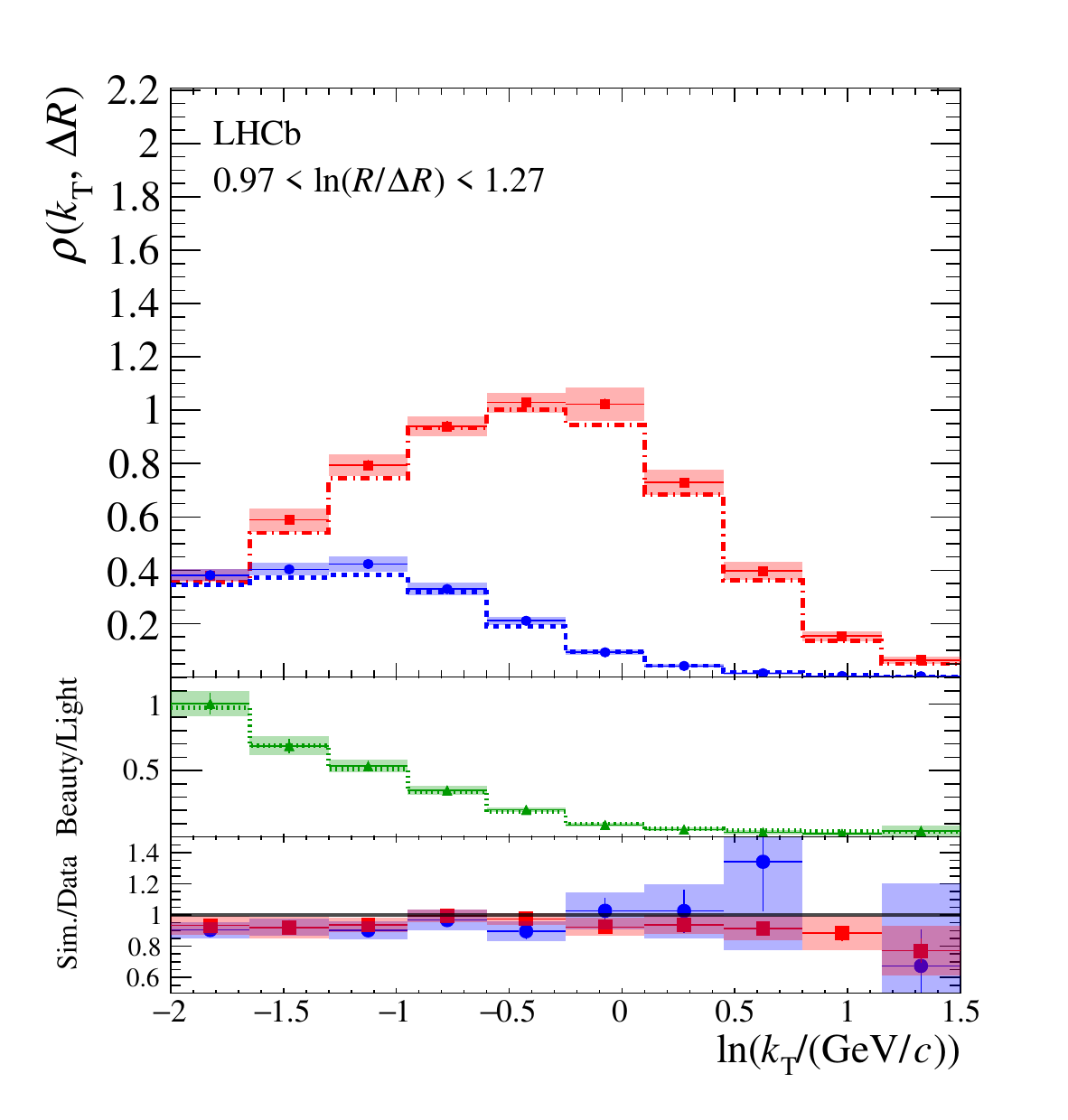}
    \end{minipage}
    \begin{minipage}[b]{0.49\textwidth}
        \includegraphics[width=\textwidth]{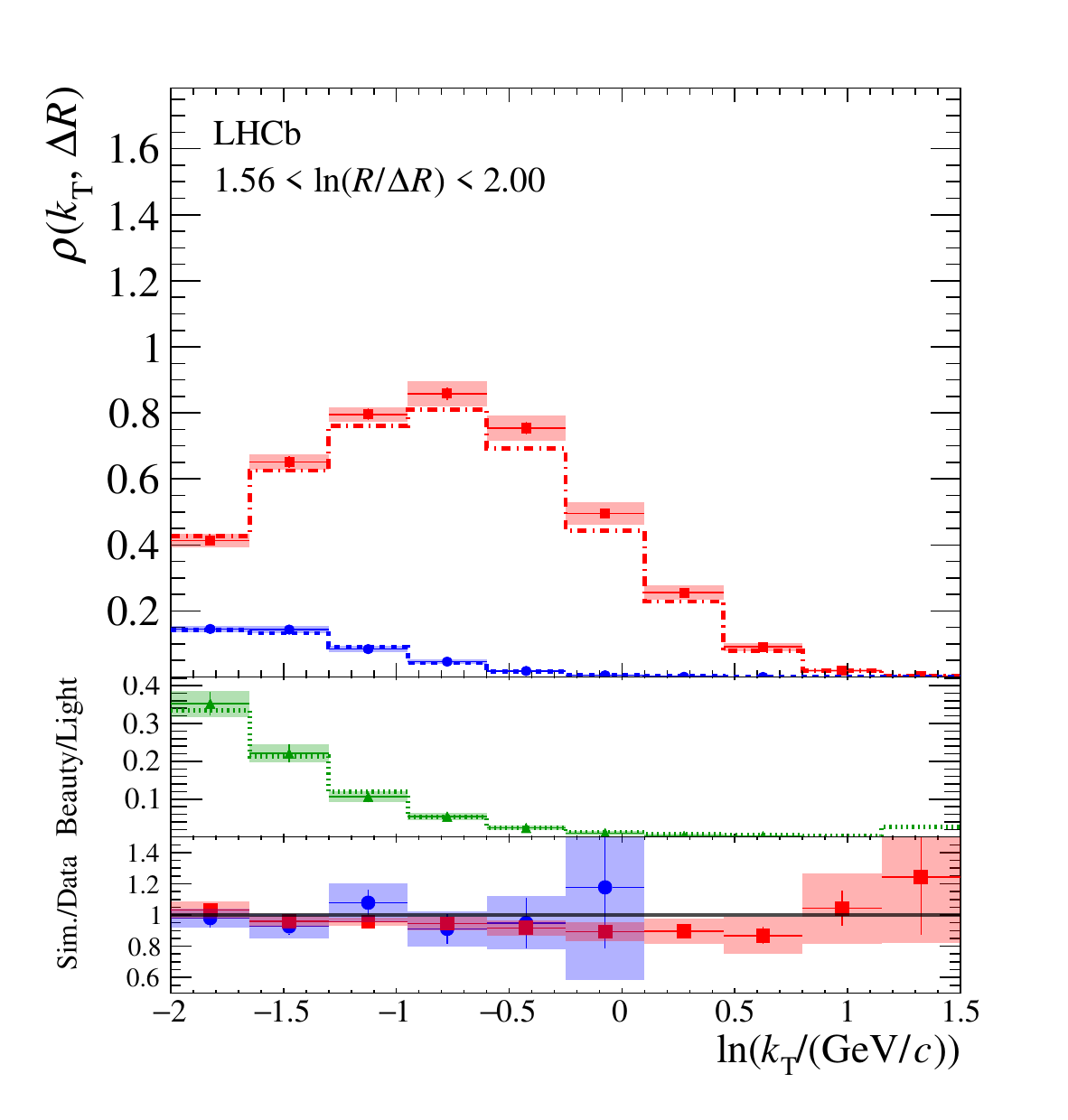}
    \end{minipage}

    \caption{Projections onto the vertical axis of the \kt-LJP for \Bpm-tagged and $Z+$jets. The ratios of beauty distributions to light-quark-enriched distributions are shown in the middle panels in data and simulation. The data are also compared with simulation in the bottom panels where the ratio of simulation to data is shown. The vertical bars represent the statistical uncertainty and the shaded areas represent the total systematic uncertainty. The projections are in bins of \lndR, starting at (top left) more wide-angle radiation $0.00 < \lndR < 0.15$ (or $0.43 < \Delta R < 0.50$) and moving towards (bottom right) more collinear radiation $1.56 < \lndR < 2.00$ (or $0.055 < \Delta R < 0.110$).
    }
    
    \label{fig:LJPktdRvertprojections}
\end{figure*}

\begin{figure*}[!h]
    \centering
    
    \begin{minipage}[b]{0.49\textwidth}
        \includegraphics[width=\textwidth]{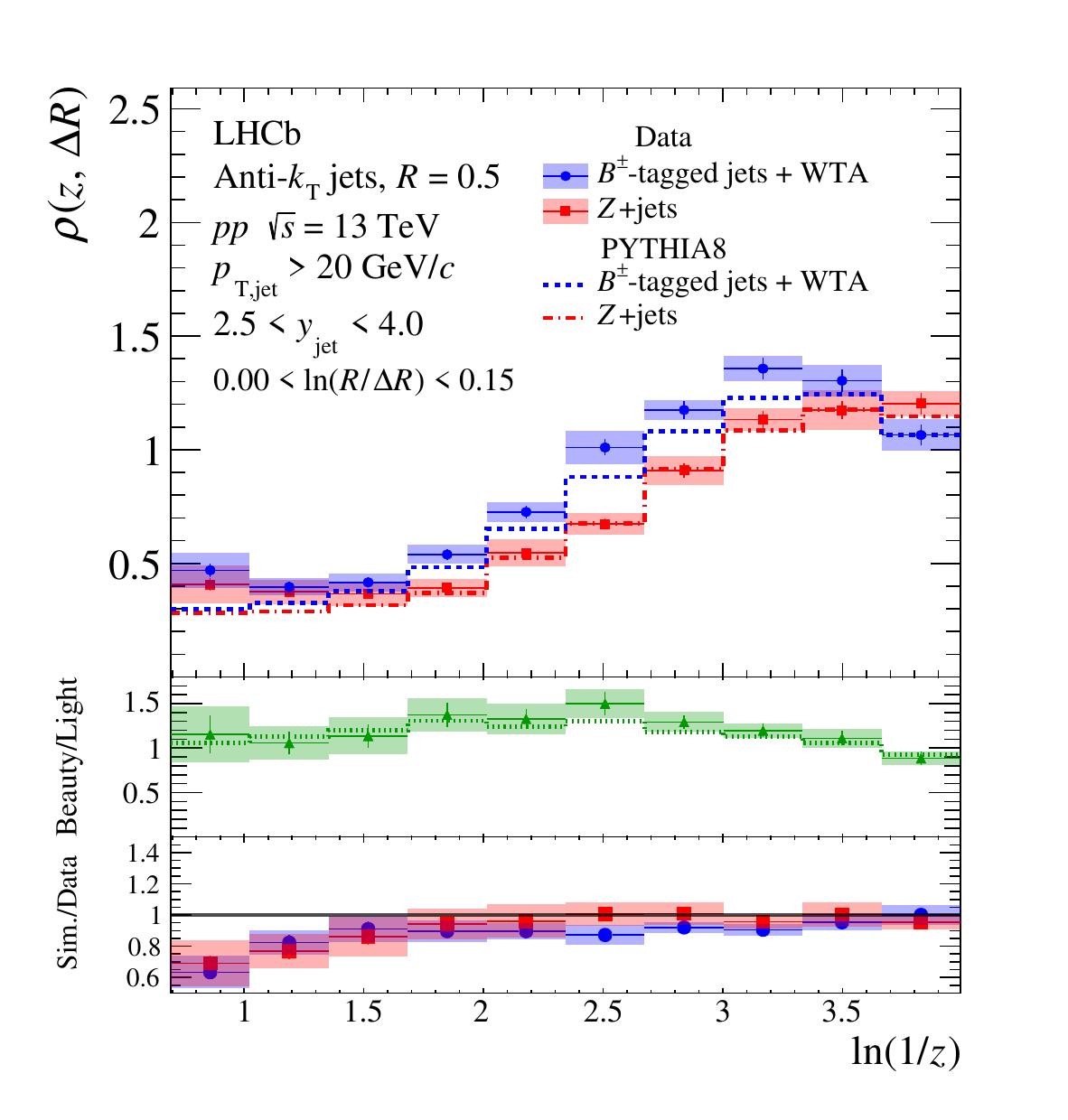}
    \end{minipage}
    \hfill
    \begin{minipage}[b]{0.49\textwidth}
        \includegraphics[width=\textwidth]{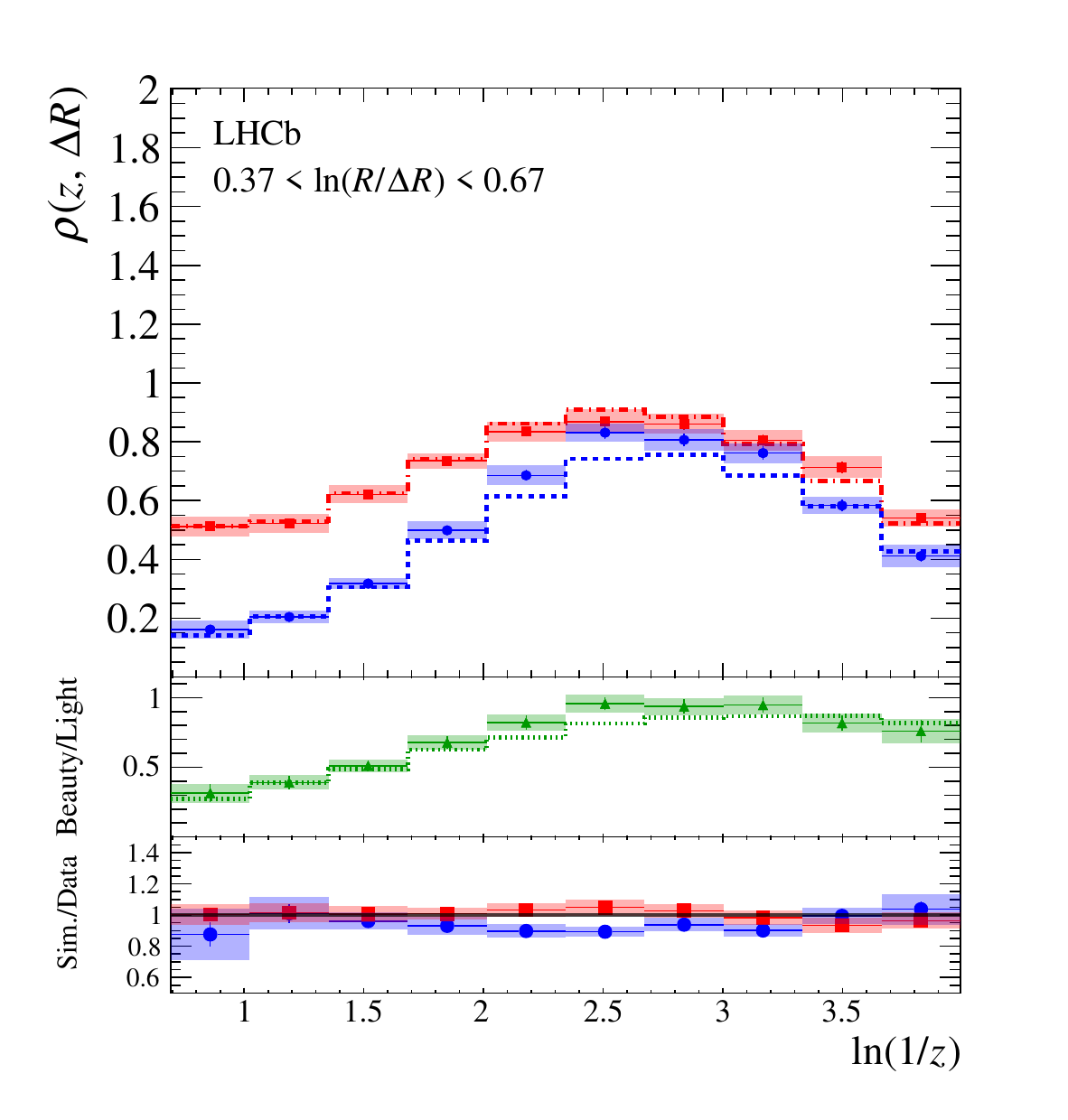}
    \end{minipage}
    \hfill
    \begin{minipage}[b]{0.49\textwidth}
        \includegraphics[width=\textwidth]{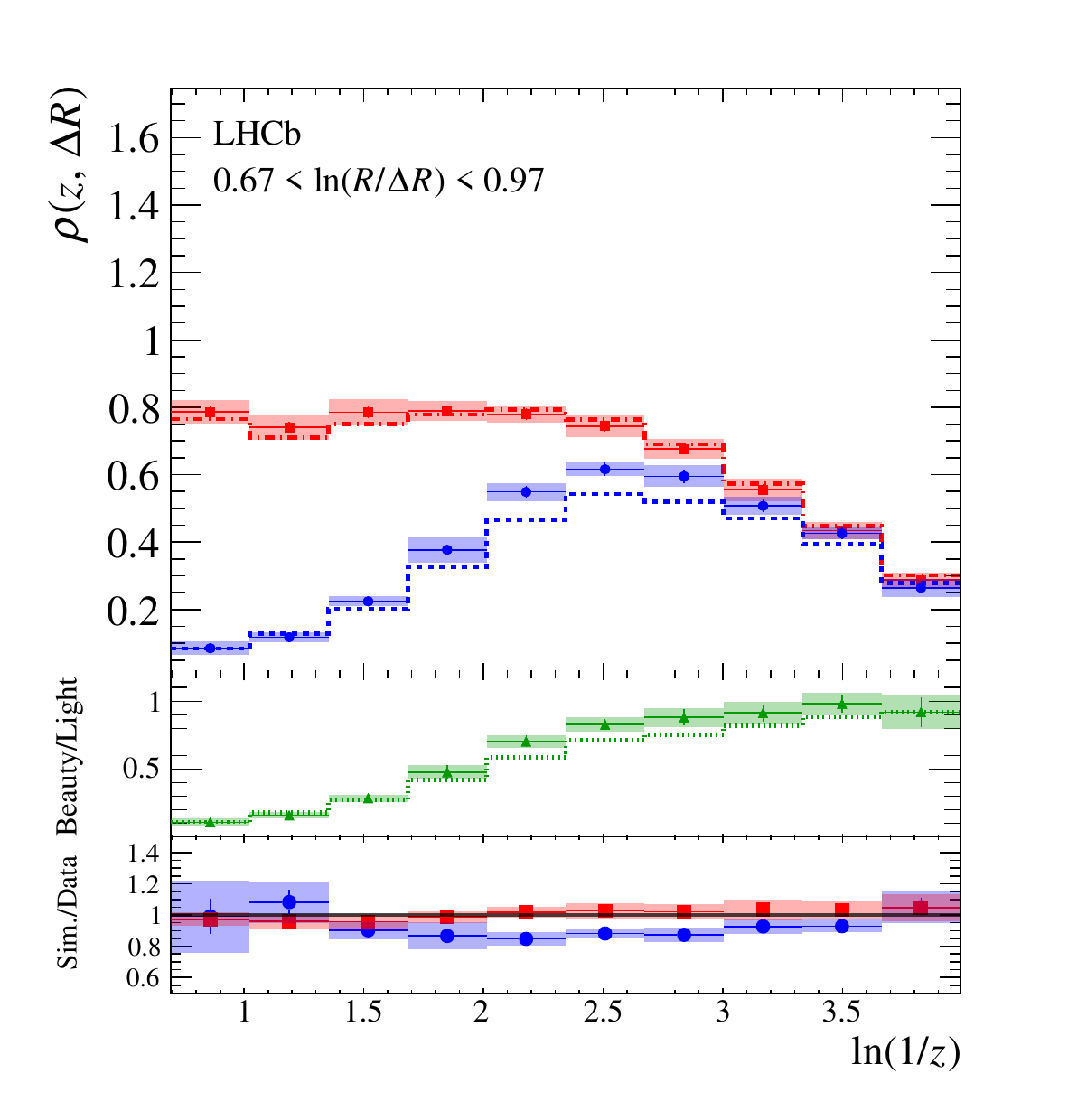}
    \end{minipage}
    \hfill
    \begin{minipage}[b]{0.49\textwidth}
        \includegraphics[width=\textwidth]{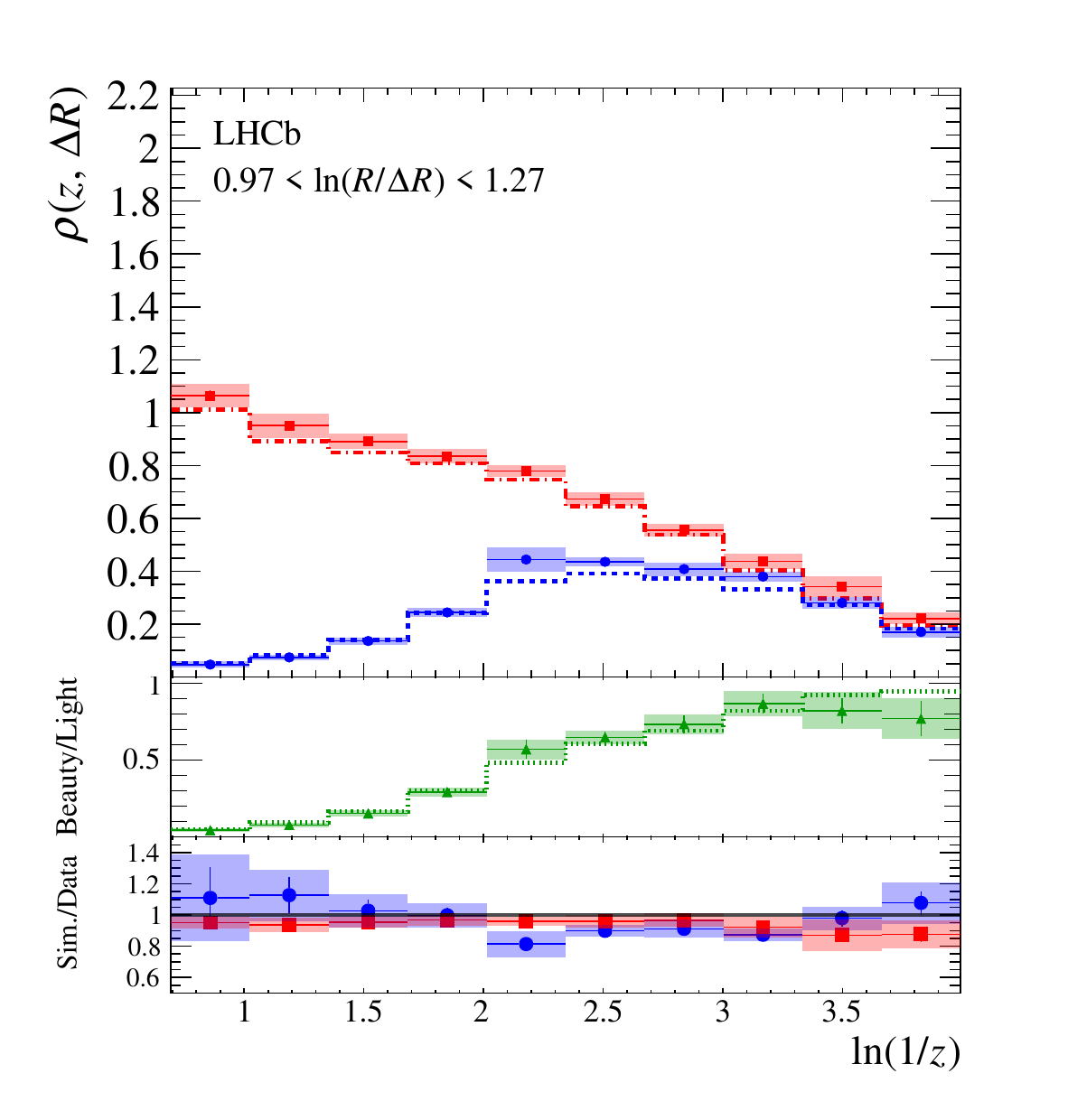}
    \end{minipage}
    \hfill

    \caption{Projections onto the vertical axis of the $z$-LJP for \Bpm-tagged and $Z+$jets. The ratios of beauty distributions to light-quark-enriched distributions are shown in the middle panels in data and simulation. The data are also compared with simulation in the bottom panels where the ratio of simulation to data is shown. The vertical bars represent the statistical uncertainty and the shaded areas represent the total systematic uncertainty. The projections are in bins of \lndR, starting at more (top left) wide-angle radiation $0.00 < \lndR < 0.15$ (or $0.43 < \Delta R < 0.50$) and moving towards (bottom right) more collinear radiation $1.56 < \lndR < 2.00$ (or $0.055 < \Delta R < 0.110$).
    }
    
    \label{fig:LJPzdRvertprojections}
\end{figure*}

\begin{figure*}[!h]
    \centering
    \begin{minipage}[b]{0.49\textwidth}
        \includegraphics[width=\textwidth]{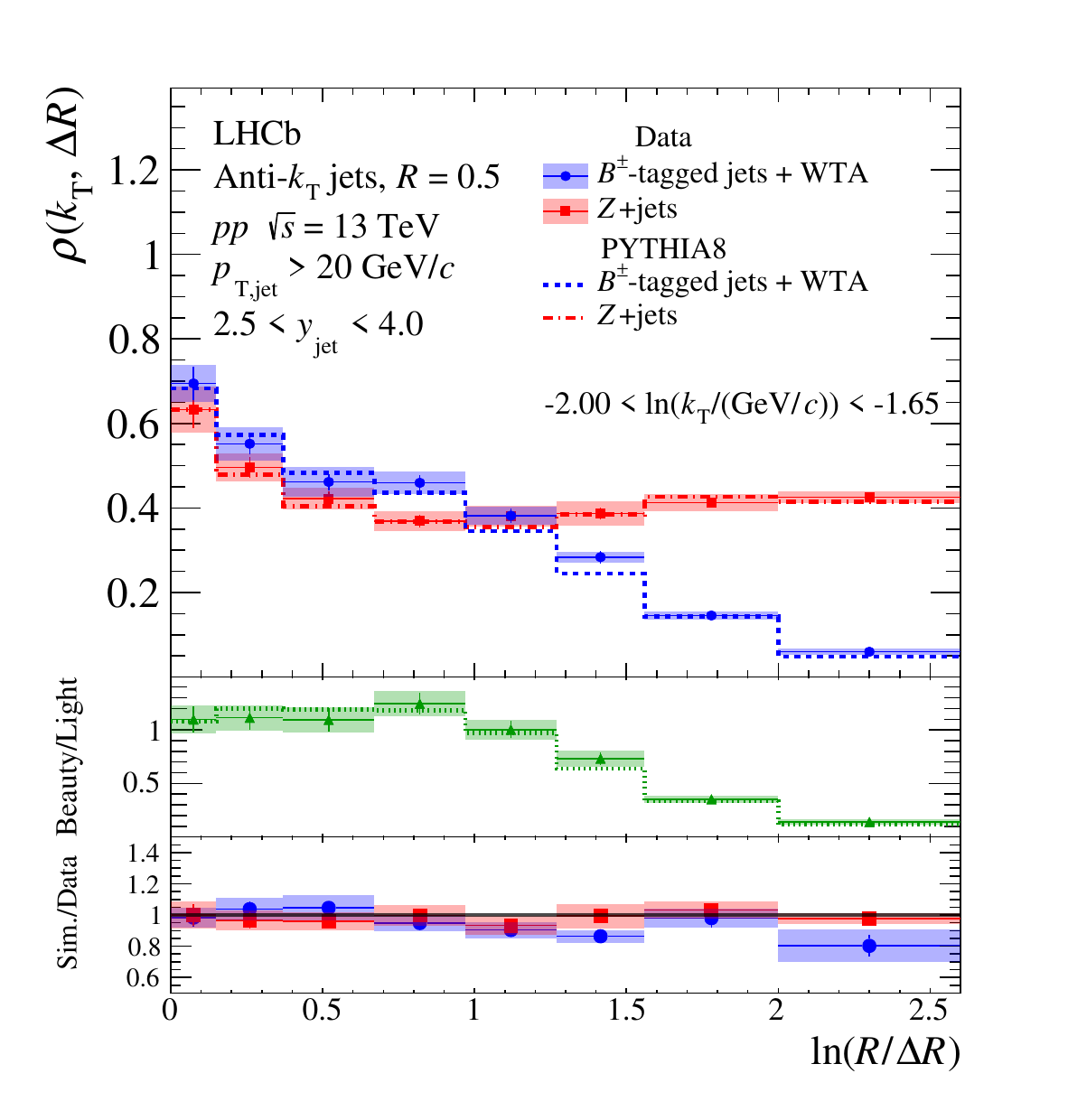}
    \end{minipage}
    \hfill
    \begin{minipage}[b]{0.49\textwidth}
        \includegraphics[width=\textwidth]{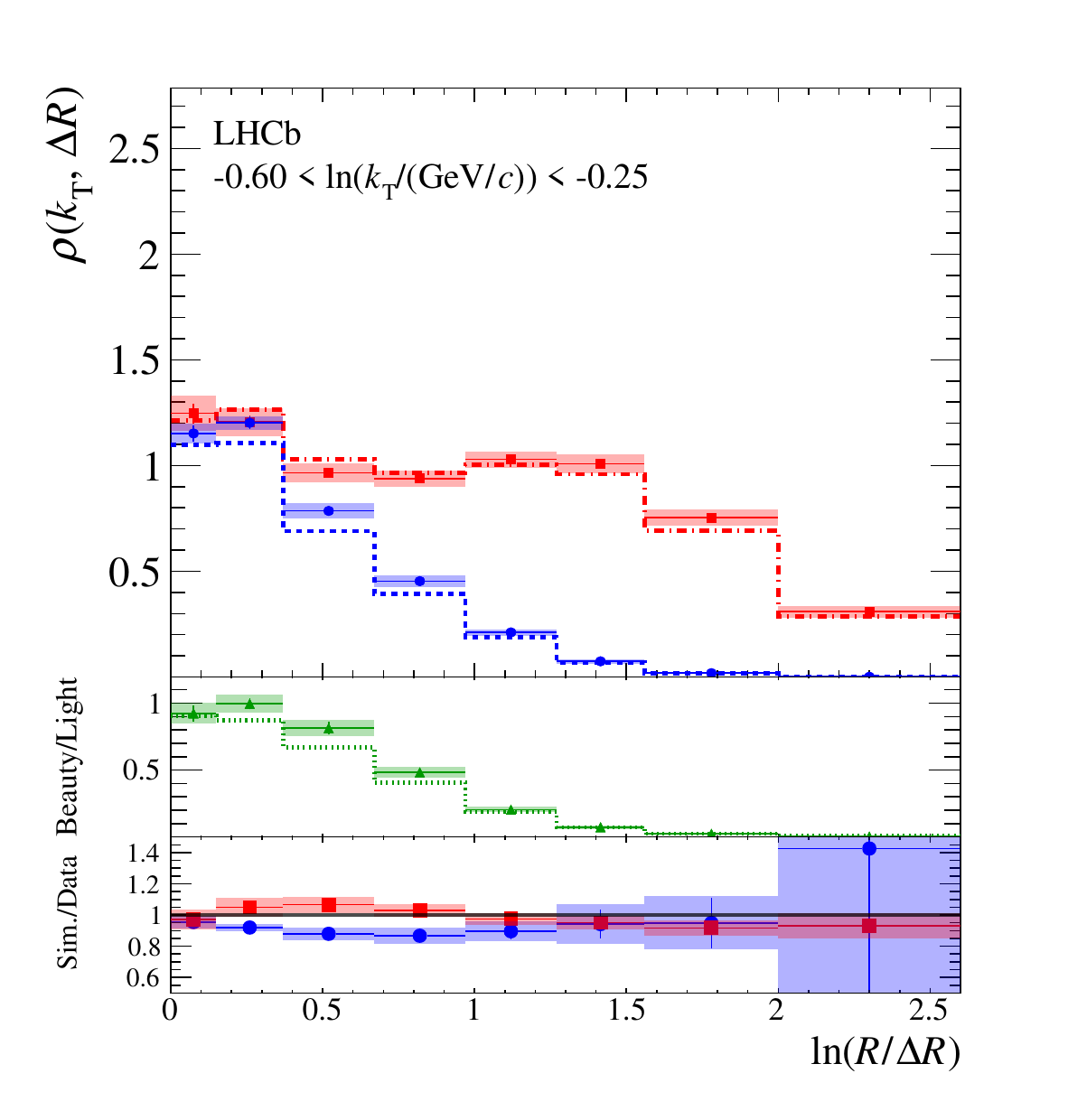}
    \end{minipage}
    \hfill
    \begin{minipage}[b]{0.49\textwidth}
        \includegraphics[width=\textwidth]{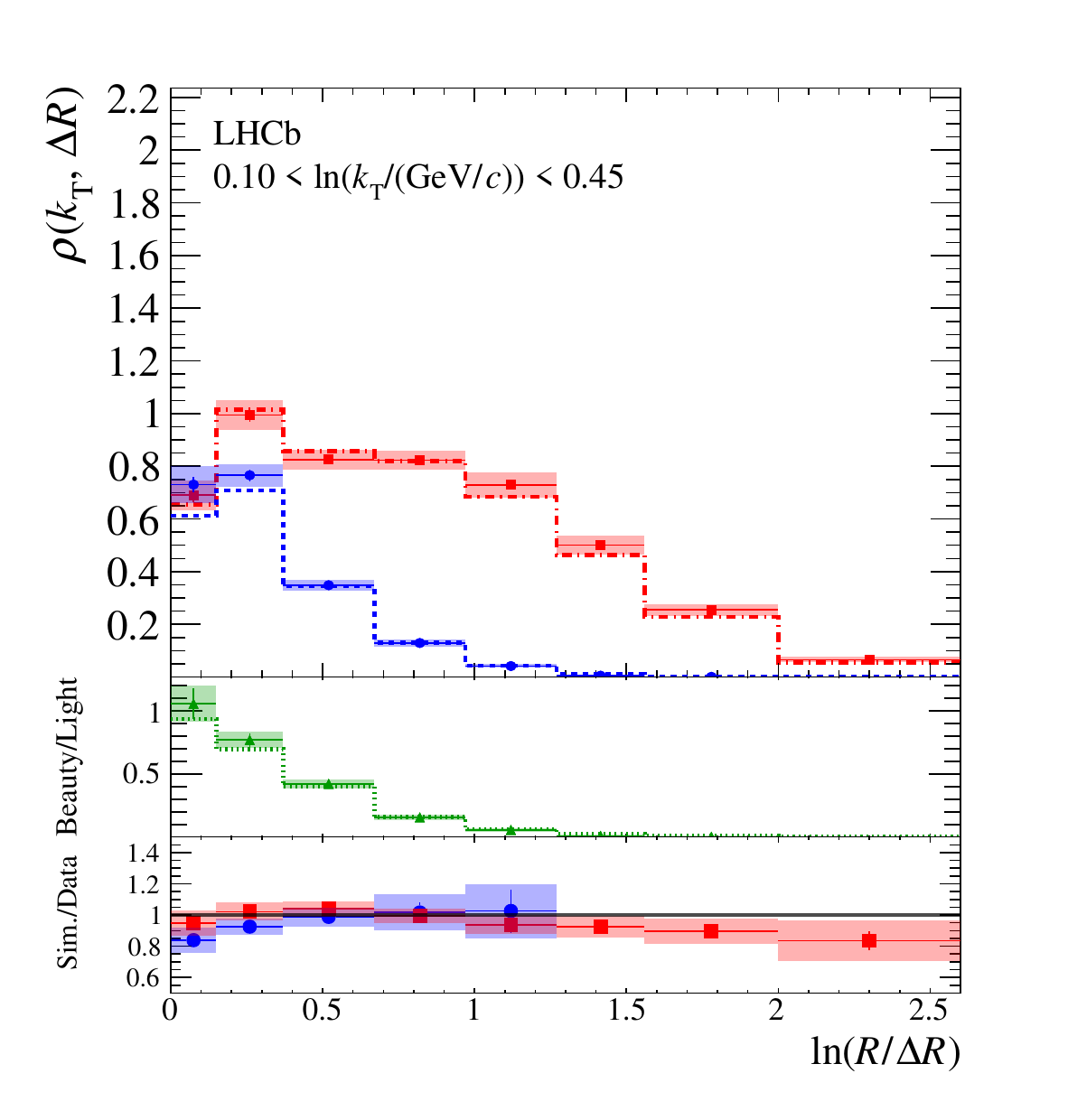}
    \end{minipage}
    \hfill
    \begin{minipage}[b]{0.49\textwidth}
        \includegraphics[width=\textwidth]{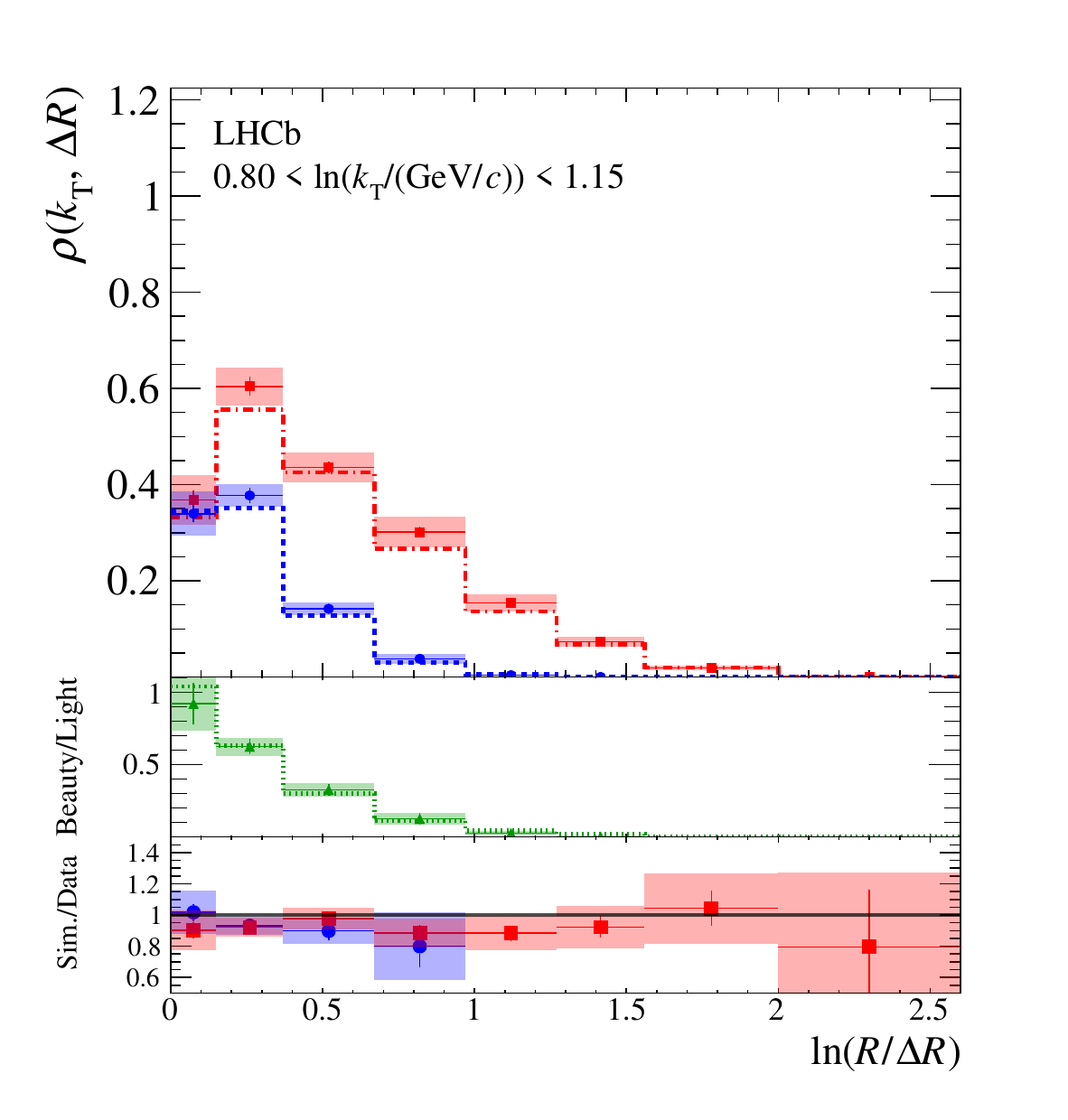}
    \end{minipage}
    \hfill

    \caption{Projections onto the horizontal axis of the \kt-LJP for \Bpm-tagged and $Z+$jets. The ratios of beauty distributions to light-quark-enriched distributions are shown in the middle panels in data and simulation. The data are also compared with simulation in the bottom panels where the ratio of simulation to data is shown. The vertical bars represent the statistical uncertainty and the shaded areas represent the total systematic uncertainty. The projections are in bins of \lnkt, starting at (top left) more nonperturbative radiation $-2.00 < \lnkt < -1.65$ (or $0.14 < k_{\mathrm{T}} < 0.19\gevc$) and moving towards (bottom right) more perturbative radiation \mbox{$0.80 < \lnkt < 1.15$} \mbox{(or $2.2 < k_{\mathrm{T}} < 3.2\gevc$)}.
    }
    
    \label{fig:LJPktdRhorprojections}
\end{figure*}

The final unfolded measurements are also shown as vertical and horizontal projections of the LJP on its variables in Figs.~\ref{fig:LJPktdRvertprojections},~\ref{fig:LJPzdRvertprojections}, and~\ref{fig:LJPktdRhorprojections}, and the data are also compared with \pythia8 predictions. Figure \ref{fig:LJPktdRvertprojections} shows projections onto the vertical axis of the \kt-LJP for $Z$+jets and \Bpm-tagged jets in data and \pythia8. The projections are shown in four bins of \lndR, starting at more wide-angle radiation $0.00 < \lndR < 0.15$ (or $0.43 < \Delta R < 0.50$) and moving towards more collinear radiation $1.56 < \lndR < 2.00$ (or $0.055 < \Delta R < 0.110$). The running of $\alpha_s(\kt)$ is observed in the distributions of $Z$+jets and \Bpm-tagged jets as described previously. The density of wide-angle radiation is virtually identical between $Z$+jets and \Bpm-tagged jets, revealing a universal behavior of quark radiation at wide angles where the dead-cone effect is negligible. The ratio of beauty to light-quark-enriched jets in the second pad shows a progressively stronger suppression of perturbative collinear radiation moving from wide angles ($0.00 < \lndR < 0.15$ or $0.43 < \Delta R < 0.50$) to narrow angles ($1.56 < \lndR < 2.00$ or $0.055 < \Delta R < 0.110$).

Similar observations can be made in Fig. \ref{fig:LJPzdRvertprojections} which shows projections onto the vertical axis of the $z$-LJP for $Z$+jets and \Bpm-tagged jets. Again, the projections are shown in four bins of \lndR, starting at more wide-angle radiation. The density of the softest emissions (large values of \lnz) remains similar between $Z$+jets and \Bpm-tagged jets across various splitting angles. On the other hand, the density of hard emissions increases in $Z$+jets with more collinear splittings, in contrast to \Bpm-tagged jets which exhibit a suppression of hard-collinear radiation. This suppression leads to the hard fragmentation of the heavy quark, maintaining most of its energy after the parton shower.

The performance of \pythia8 on $Z+$jets is good, with a simulation-to-data ratio consistent with unity across most of the LJP projections. For \Bpm-tagged jets, \pythia8 slightly underpredicts the emission density up to 25\% in the bulk of the LJP, as seen in Figs. \ref{fig:LJPktdRvertprojections} and \ref{fig:LJPzdRvertprojections}.
A discrepancy up to 30--40\% between \pythia8 and data is observed in the hard and wide-angle splittings (low \lnz region in the top left plot) for both $Z$+jets and \Bpm-tagged jets. 

Figure \ref{fig:LJPktdRhorprojections} shows projections onto the horizontal axis of the \kt-LJP for $Z$+jets and \Bpm-tagged jets in data and \pythia8. The projections are shown in four bins of \lnkt, starting at more nonperturbative radiation \mbox{$-2.00 < \lnkt < -1.65$} \mbox{(or $0.14 < k_{\mathrm{T}} < 0.19\gevc$)} and moving towards more perturbative radiation $0.80 < \lnkt < 1.15$ (or \mbox{$2.2 < k_{\mathrm{T}} < 3.2\gevc$}). The distributions reveal a progressively stronger depletion of emissions in the collinear region (large values of \lndR) for \Bpm-tagged jets relative to $Z+$jets. This is the result of the dead-cone effect in heavy-quark jets. 

\clearpage
\section{Summary}
This article presented the first measurement of the LJP of light-quark-enriched jets and beauty-quark jets. The measurement is performed on  $pp$ collision data collected by the \lhcb experiment at $\sqrt{s} = 13\tev$, corresponding to an integrated luminosity of 5.4\invfb. Jets are reconstructed with a particle flow algorithm including charged and neutral candidates, and the clustering is performed using the anti-\kt algorithm with $R = 0.5$ and reclustered using the C/A algorithm to exploit the angular ordering of QCD radiation. The selected jets are required to have $\pt > 20\gevc$ and $2.5 < y < 4.0$ to ensure the full jet is within the \lhcb acceptance and that jet constituents are well-reconstructed. Jets that are back-to-back with a $Z$ boson are used to obtain a light-quark-enriched jet sample, and jets built around fully reconstructed \Bpm mesons tagged with WTA flavor are used to obtain a beauty-jet sample. The detector inefficiencies and bin migrations are studied in simulation using \pythia8, which is then used to correct and unfold the measured distributions to particle level. The measurement reveals the phase space of QCD splittings for light and beauty quarks. Stark differences are seen between the emission densities of $Z+$jets and \Bpm-tagged jets, where the latter exhibits a severe suppression of radiation in the collinear region of the LJP relative to the former. This dead-cone effect in beauty jets suppresses hard collinear radiation off of beauty quarks, which results in the beauty quark maintaining most of its energy. This is why beauty quarks exhibit hard fragmentation. The data will serve as a precision test of QCD in future comparisons with theoretical calculations.




\section*{Acknowledgements}
%
%
\noindent We express our gratitude to our colleagues in the CERN
accelerator departments for the excellent performance of the LHC. We
thank the technical and administrative staff at the LHCb
institutes.
We acknowledge support from CERN and from the national agencies:
ARC (Australia);
CAPES, CNPq, FAPERJ and FINEP (Brazil); 
MOST and NSFC (China); 
CNRS/IN2P3 (France); 
BMBF, DFG and MPG (Germany); 
INFN (Italy); 
NWO (Netherlands); 
MNiSW and NCN (Poland); 
MCID/IFA (Romania); 
MICIU and AEI (Spain);
SNSF and SER (Switzerland); 
NASU (Ukraine); 
STFC (United Kingdom); 
DOE NP and NSF (USA).
We acknowledge the computing resources that are provided by ARDC (Australia), 
CBPF (Brazil),
CERN, 
IHEP and LZU (China),
IN2P3 (France), 
KIT and DESY (Germany), 
INFN (Italy), 
SURF (Netherlands),
Polish WLCG (Poland),
IFIN-HH (Romania), 
PIC (Spain), CSCS (Switzerland), 
and GridPP (United Kingdom).
We are indebted to the communities behind the multiple open-source
software packages on which we depend.
Individual groups or members have received support from
Key Research Program of Frontier Sciences of CAS, CAS PIFI, CAS CCEPP, 
Fundamental Research Funds for the Central Universities,  and Sci.\ \& Tech.\ Program of Guangzhou (China);
Minciencias (Colombia);
EPLANET, Marie Sk\l{}odowska-Curie Actions, ERC and NextGenerationEU (European Union);
A*MIDEX, ANR, IPhU and Labex P2IO, and R\'{e}gion Auvergne-Rh\^{o}ne-Alpes (France);
Alexander-von-Humboldt Foundation (Germany);
ICSC (Italy); 
Severo Ochoa and Mar\'ia de Maeztu Units of Excellence, GVA, XuntaGal, GENCAT, InTalent-Inditex and Prog.~Atracci\'on Talento CM (Spain);
SRC (Sweden);
the Leverhulme Trust, the Royal Society and UKRI (United Kingdom).


\newpage
\appendix


\addcontentsline{toc}{section}{References}
\bibliographystyle{LHCb}
\bibliography{main,standard,LHCb-PAPER,LHCb-TDR,LHCb-DP}

\newpage
\centerline
{\large\bf LHCb collaboration}
\begin
{flushleft}
\small
R.~Aaij$^{38}$\lhcborcid{0000-0003-0533-1952},
A.S.W.~Abdelmotteleb$^{57}$\lhcborcid{0000-0001-7905-0542},
C.~Abellan~Beteta$^{51}$\lhcborcid{0009-0009-0869-6798},
F.~Abudin{\'e}n$^{57}$\lhcborcid{0000-0002-6737-3528},
T.~Ackernley$^{61}$\lhcborcid{0000-0002-5951-3498},
A. A. ~Adefisoye$^{69}$\lhcborcid{0000-0003-2448-1550},
B.~Adeva$^{47}$\lhcborcid{0000-0001-9756-3712},
M.~Adinolfi$^{55}$\lhcborcid{0000-0002-1326-1264},
P.~Adlarson$^{84}$\lhcborcid{0000-0001-6280-3851},
C.~Agapopoulou$^{14}$\lhcborcid{0000-0002-2368-0147}
% ...

{\footnotesize \it

% ...
$^{14}$Universit{\'e} Paris-Saclay, CNRS/IN2P3, IJCLab, Orsay, France\\
% ...
$^{38}$Nikhef National Institute for Subatomic Physics, Amsterdam, Netherlands\\
% ...
$^{47}$Instituto Galego de F{\'\i}sica de Altas Enerx{\'\i}as (IGFAE), Universidade de Santiago de Compostela, Santiago de Compostela, Spain\\
% ...
$^{51}$Physik-Institut, Universit{\"a}t Z{\"u}rich, Z{\"u}rich, Switzerland\\
% ...
$^{55}$H.H. Wills Physics Laboratory, University of Bristol, Bristol, United Kingdom\\
% ...
$^{57}$Department of Physics, University of Warwick, Coventry, United Kingdom\\
% ...
$^{60}$School of Physics and Astronomy, University of Glasgow, Glasgow, United Kingdom\\
$^{61}$Oliver Lodge Laboratory, University of Liverpool, Liverpool, United Kingdom\\
% ...
$^{69}$Syracuse University, Syracuse, NY, United States\\
% ...
$^{84}$Department of Physics and Astronomy, Uppsala University, Uppsala, Sweden, associated to $^{60}$\\
% ...
}
\end{flushleft}


\end{document}